
\input phyzzx.tex
%
\catcode`\@=11 
\def\papers{\papersize\headline=\paperheadline\footline=\paperfootline}
\def\papersize{\hsize=40pc \vsize=53pc \hoffset=0pc \voffset=1pc
   \advance\hoffset by\HOFFSET \advance\voffset by\VOFFSET
   \pagebottomfiller=0pc
   \skip\footins=\bigskipamount \normalspace }
\catcode`\@=12 
\papers
%
\def\HG{{\widehat \Gamma}}
\def\H{{\cal H}}
\def\d{\delta}
\def\O{{\cal O}}
\def\T{{\cal T}}

\def\e{{\epsilon}}
\def\D{{\cal D}}
\def\ket#1{\mid #1 \rangle}
\def\bra#1{\langle#1 \mid}
\def\vev#1{\left\langle #1 \right\rangle}
\def\VEV#1{\bigl\langle #1 \bigr\rangle}
\overfullrule=0pt
\baselineskip 13pt
\pubnum{MIT-CTP-2193}
\date{April 1993}
\titlepage
\title{CONNECTIONS ON THE STATE-SPACE\break
OVER CONFORMAL FIELD THEORIES}
\author{K. Ranganathan\foot{Supported in part
D.O.E. contract DE-AC02-76ER03069 and NSF grant PHY91-06210.},
H. Sonoda\foot{Permanent address: Department of Physics,
UCLA, Los Angeles, CA 90024-1547, USA. \hfill \break
Supported in part by D.O.E. contract
DE-AT03-88ER 40384 Mod A006 Task C.},
and B. Zwiebach${}^\star$}
\address{Center for Theoretical Physics \break
Laboratory for Nuclear Science \break
and Department of Physics \break
Massachusetts Institute of Technology \break
Cambridge, MA 02139, USA}

\abstract{Motivated by the problem of background independence
of closed string field theory we study geometry on the infinite
vector bundle of local fields over the space of conformal field
theories (CFT's). With any connection we can associate an
excluded domain $D$ for the integral of marginal operators, and
an operator one-form $\omega_\mu$. The pair $(D, \omega_\mu)$
determines the covariant derivative of any correlator of local fields.
We obtain interesting classes of connections
in which $\omega_\mu$'s can be written in terms of CFT data. For these
connections we compute their curvatures in terms of
four-point correlators, $D$, and $\omega_\mu$. Among these connections three
are of particular interest. A flat, metric
compatible connection $\HG$, and connections $c$ and $\bar c$ with
non-vanishing curvature, with the latter metric compatible. The flat
connection cannot be used to do parallel transport over a finite
distance. Parallel transport with either $c$ or $\bar c$, however,
allows us to construct a CFT in the state space of another CFT
a finite distance away. The construction is given in the form of
perturbation theory manifestly free of divergences.}

\endpage

\REF\rfriedanplus{D. Friedan, Ann. Phys.{\bf 163}(1985) 318; \hfill\break
C. G. Callan, D. Friedan, E. J. Martinec and M. J. Perry,
Nucl. Phys. {\bf B280} (1987) 599; \hfill\break
A. Sen, Phys. Rev. Lett. {\bf 55} (1985) 1846;\hfill\break
C .Lovelace, Phys. Lett. {\bf B135} (1984) 75;\hfill\break
E. S. Fradkin and A. A. Tseytlin, Phys. Lett. {\bf B158} (1985) 316;
{\bf B160} (1985) 69.}
\REF\zwiebachlong{B. Zwiebach, ``Closed string field theory:
quantum action and the Batalin-Vilkovisky master equation,''
Nucl. Phys. {\bf B390} (1993) 33.}
\REF\rsen{A. Sen, Nucl. Phys. {\bf B345} (1990) 551; {\bf B347}
(1990) 270.}
\REF\rwitten{E. Witten, Phys. Rev. {\bf D46} (1992) 5446.}
\REF\rwittendev{E. Witten, ``Some calculations
in background independent off-shell
string theory, IAS preprint, hep-th/9210065;\hfill\break
K. Li and E. Witten, ``Role of Short Distance Behavior in Off-Shell
Open String Field Theory'', IASSNS-HEP-93/7 ;\hfill\break
S. L. Shatashvili, ``Comment on the Background Independent Open String
Theory'', IASSNS-HEP-93/15.}
\REF\rschwarz{A. Schwarz, ``Geometry of Batalin-Vilkovisky
Quantization'', UC Davis preprint, hep-th/9210115 .}
\REF\rhatazw{H. Hata and B. Zwiebach, ``Developing the Covariant
Batalin-Vilkovisky Approach to String Theory'', MIT-preprint,
hep-th/9301097, to appear in Ann. Phys.}
\REF\rkz{T. Kugo and B. Zwiebach, Prog. Theo. Phys. {\bf 87} (1992) 801}
\REF\rsonoda{H. Sonoda, Nucl. Phys. {\bf B383} (1992) 173;\hfill\break
``Operator Coefficients for Composite Operators in the $(\phi^4)_4$
Theory'',  to appear in Nucl. Phys. B}
\REF\rranga{K. Ranganathan, ``Nearby CFT's in the operator formalism:
The role of a connection,''  MIT-CTP-2154, hep-th/9210090.}
\REF\rcnw{M. Campbell, P. Nelson, and E. Wong, Int. J. Mod. Phys.
{\bf A6} (1991) 4909.}
\REF\rkutasov{D. Kutasov, Phys. Lett. {\bf B220} (1989) 153.}
\REF\rsegal{G. Segal, ``The definition of Conformal Field Theory,"
Oxford preprint.}
\REF\rcecottivafa {S. Cecotti and C. Vafa, ``Topological and
Anti-topological fusion'', HUTP-91/A031;\hfill\break
S. Cecotti, Int. J. Mod. Phys. {\bf A6} (1991) 1749;\hfill\break
S. Cecotti, Nucl. Phys. {\bf B355} (1991) 755.}
\REF\rdijk{L. Kadanoff, Ann. Phys. {\bf 120} (1979) 39;\hfill\break
J. Cardy, Jour. Phys. A. (1987)  L891;\hfill\break
R. Dijkgraaf, E. Verlinde and H. Verlinde, in ``Perspectives of
String Theory'', ed. P. Di Vecchia and J. L. Petersen (World Scientific,
Singapore, 1988);\hfill\break
S. Chaudhuri and J. A. Schwartz, Phys. Lett. {\bf B219} (1989) 291. }

\chapter{Introduction and Summary}

It has been speculated for some time that a background independent
and geometrical formulation of closed string field theory must use
the space of all possible two dimensional field theories.
Though this theory space has not yet been constructed, one would
expect the string action to be a function over the theory space,
and conformal field theories to appear as local extrema of the
action, namely, as classical solutions to the string field equations.
This intuition is backed by the work that showed that the spacetime
string equations of motion for massless fields can be derived from
the conditions of conformal invariance of two dimensional theories
[\rfriedanplus ].
Presently we only know how to formulate
closed string field theory given a particular conformal field
theory (see, for example, [\zwiebachlong].)
The action is a function over the tangent space
to the theory space at the conformal field theory, since
each possible theory deformation corresponds to a particle
field in space-time. Background independence demands
that closed string field theories formulated
with different conformal field theories must be physically equivalent.
This question, in the case of nearby theories, has been considered in
refs.~[\rsen]. More recently, it has been suggested [\rwitten ] that the
Batalin-Vilkovisky formulation of string field theory gives us
strong indications of the geometrical structures expected to
exist in theory space (see also [\rwittendev ,\rschwarz ,\rhatazw ]).
The possible
relevance of two-dimensional theory space to the formulation
of closed string theory was a major motivation for the present
investigation.

When we have conformal field theories with continuous parameters
$x^\mu$, we must make sure that the closed string field theory
formulated at each $x^\mu$ is equivalent to one another.
(In the case of toroidal compactification, the constant target
space metric and antisymmetric tensor provide natural coordinates
$x^\mu$.) Closed string field theory  [\zwiebachlong]
uses the operator formalism for conformal
field theory, in which a ket-state $\ket{\Phi_i}_x$ is introduced
for each local field $\Phi_i$ on the world sheet.
We call the space of all ket-states $\H_x$ and the dual space
of all bra-states $\H_x^*$.  At each $x^\mu$
there is a distinct space $\H_x$, and the $\H$'s form an infinite
dimensional vector bundle over the space of conformal field
theories with coordinates $x^\mu$.
Hence, to address the question of background independence properly,
we must compare two linear spaces $\H_x$ and $\H_{x+\d x}$.
This requires the introduction of a connection.
In the case of toroidal compactification a connection
$\Gamma_\mu$ has been introduced and constructed explicitly
in ref.~[\rkz].

This work was motivated by two earlier works, each of which discussed
the role of a connection. In one
work [\rsonoda] a connection on theory space has been obtained
by working with the space of renormalized field theories
parametrized by $x^\mu$. The correlators of local fields depend
on $x$'s smoothly and the partial derivative of a correlator with
respect to $x^\mu$ is given in terms of a spatial integral of a
correlator with an additional insertion of the operator $\O_\mu$
conjugate to $x^\mu$. The integral diverges due to the short distance
singuarities between $\O_\mu$ and local fields.  Naively we may simply
regulate the integral by restricting the range of integration
and take the limit after subtracting the divergences.  But,
in general, we need extra finite counterterms.
The coefficients of these finite counterterms can be interpreted
as a connection for the vector bundle of local fields over
the theory space parametrized by $x$'s.  In refs.~[\rsonoda]
the formula expressing the covariant derivatives of correlators
has been called the variational formula.

In another work [\rranga] the operator formulation of CFT's
was used to introduce a connection. In the operator formalism,
a conformal field theory at $x$ assigns to each punctured Riemann
surface $\Sigma$ a tensor $\bra{\Sigma}$ on $\H_x^*$. Therefore,
each surface $\Sigma$ defines a section ${}_x\bra{\Sigma}$ over
theory space. Such sections will be called surface sections.
In ref.~[\rcnw] sewing was used to single out a deformation
of the states $\bra{\Sigma}$.
The deformed surface states, even though they represent a
different theory, were defined using the original state space $\H_x$.
A prescription to compare the deformed states with
those at $\H_{x+\d x}$ (for toroidal compactification) has been
given recently [\rranga]. It was established in that work that
parallel transport with the connection of ref.~[\rkz] precisely
takes the surface states at $\H_{x+\d x}$ into
the deformed surface states of ref.~[\rcnw].

\subsection{The present work.}
In the present paper theory space will be a space of conformal
field theories.  The vector space $\H_x$ at each point $x$ is
infinite dimensional and will be spanned by an infinite
set of basis states. Our aim is to investegate the connections
in this vector bundle. We now summarize our main results.

If we consider our vector bundle without regard for the CFT structure
then all connections are on an equal footing. The existence of the
surface sections $\bra{\Sigma}$ which encode the CFT structures allows
us to distinguish covariant derivatives based on their action on these
sections. A connection is natural if the covariant derivatives of
the surface sections can be given in terms of intrinsic operations
in the conformal theories. We will now explain this idea.

As a first step we recognize a very general fact. For any connection
$\Gamma_\mu$ we show that the associated covariant derivative
$D_\mu (\Gamma)$ always generates a CFT deformation. That is, the
deformed surface states
$\bra{\Sigma} + \delta x^\mu D_\mu (\Gamma) \bra{\Sigma}$ satisfy
the same sewing relations as the original surface states $\bra{\Sigma}$,
and therefore, define a conformal field theory.

As a second step we break the space of CFT deformations into two
types. In the first type we deform the surface states by integrating
the insertion of marginal operators over the surfaces minus some
disks. This deformation changes the CFT. The second type of deformation
is trivial (but essential). A similarity transformation generated
by an arbitrary operator $\omega_\mu$ is
performed on every surface state. This does not change the CFT.
Since infinitesimal deformations form a vector space, these
two types of transformations can be added. The most
general deformations that we will have to consider will be arbitrary
linear combinations of these two types of deformations.

The above two steps lead us to expect that the deformations generated by
$\Gamma$ can be written in the above form. In fact it is appropriate to regard
as the very definition of a family of CFT's the requirement that for any
connection $\Gamma$ there is an $\omega_\mu$ and $\O_\mu$
such that $D_\mu (\Gamma)$ of surface sections can be written as
$$D_\mu (\Gamma) \bra{\Sigma} = - \hskip-6pt \int_{\Sigma -
\cup_i D_i}\hskip-6pt d^2 z~ \bra{\Sigma;z} \O_\mu \rangle
-  \sum_{i=1}^n \bra{\Sigma}\omega_\mu^{(i)}.\eqn\xegen$$
It will become clear later that this is an appropriate way to formalize earlier
intuitions on deforming correlation functions using the marginal
operators [\rdijk].
This equation is the starting point of much of our investigation.
It associates with any connection $\Gamma$ a pair $(D, \omega_\mu)$.
This is ambiguous, however, since many different pairs are equivalent
in that they yield the same right hand side in Eqn.\xegen . First, if $D$ is
changed, $\omega_\mu$ can be changed to yield the same right hand side.
Second, for fixed $D$, we can add to $\omega_\mu$ a symmetry one form
$S_\mu$ (a Kac-Moody symmetry on every CFT of the space is a
typical case) and again the right hand side of \xegen\ is unchanged. Keeping
these ambiguities in mind we speak of a pair $(D, \omega_\mu)$
associated with a connection and this notion will be central to
the developments in this paper.

We had stated that we would distinguish some connections. To do
so we recognize that in the first term of \xegen\ the marginal
operator is integrated over the surface minus some domains
surrounding the punctures. The domain $D$ on each puncture must
be identical in terms of the local coordinates chosen at the punctures.
This is a natural CFT operation defined all over
theory space without the need of further data.\foot{The specification
of the local coordinates around the punctures is part of the specification
of the surfaces $\Sigma$ and is not CFT data.} This allows us
to distinguish a class of connections $\HG_D$, labelled by the
excluded domain $D$ as those which are associated
with the pairs $(D, 0)$.  For $D$ the complete unit disk, we obtain
the connection $\HG$ which was explored for the case of toroidal
backgrounds [\rkz ,\rranga]. Another natural possibility for
$\omega_\mu$ will be obtained by letting the domains go to zero
and subtracting away divergences. In this case $\omega_\mu$ is
chosen to be the divergent part of an operator that arises from
three point functions. This will lead to the connection $c$ of
ref.~[\rsonoda]. With the same domain, but with some modification
of $\omega_\mu$ we are led to a related connection $\bar c$.

We investigate how to compute the connection coefficients
$\Gamma_{\mu i}^{~j}$ corresponding to a given pair
$(D, \omega_\mu)$. We may view \xegen\ as an infinite set of
linear equations for the connection coefficients - one equation
for each surface $\Sigma$. As we said earlier, corresponding to
any pair $(D, \omega_\mu)$ there are many connections differing
from each other by symmetry one forms.
Despite the ambiguity due to symmetries we develop techniques to
solve for most of the connection coefficients in terms of operator
product expansion (OPE) coefficients, and $\omega_\mu$.
We will do so by considering the equations obtained by taking
the covariant derivatives of the Virasoro operators (these operators
can be expressed in terms of surface states). We will find that the
connection coefficients $\Gamma_{\mu i}^{~~j}$ for
$\gamma_i \neq \gamma_j$ can be expressed in terms of the
OPE coefficients and $\omega_\mu$. Further, for $\gamma_i = \gamma_j$,
all the connection coefficients, $\Gamma_{\mu i}^{~~j}$, can be expressed
in terms of connection coefficients relating primary fields of
equal dimensionality. We will also discuss a way of computing
the connection coefficients using symmetry currents $J_a(z)$.

We next discuss the computation of curvature of a connection $\Gamma$
in terms of a pair $(D, \omega_\mu)$ and the states $\bra{\Sigma}$
of the theory. We cannot obtain such an expression for an arbitrary
connection. If this were possible then all connections
related by a symmetry one form will have curvatures
related by a symmetry. This we will show is false. However for the
interesting connections that we listed earlier we can obtain such
an expression by virtue of the special properties of their $\omega_\mu$'s.
For the connection $c$ our expression for curvature is algebraic in
terms of OPE coefficients. We also prove that $\HG_D$ is flat.
In the computation of the curvature of $c$ we find a consistency
condition that must be satisfied. Unless
$c_{\mu\nu}^{~~\rho} -  c_{\nu\mu}^{~~\rho}=0$, the commutator
of two covariant derivatives would give, in addition to curvature, a
term corresponding to torsion.  On a vector
bundle, however, there cannot be torsion. We know that $c_{\mu\nu}^{~~\rho} -
c_{\nu\mu}^{~~\rho}=0$  in particular
examples, and believe that this requirement follows in general from the earlier
requirement that the right hand side of \xegen\ spans the space of
deformations.

We study finite distance parallel transport with the
connections. In finite dimensional vector bundles this
is always possible. This becomes subtle in the infinite dimensional vector
bundles we are working with. We  find that the flat
connection $\HG$ diverges in second order in the perturbation
expansion in terms of the distance. The connection $c$, which
can be shown to be upper triangular, leads to perturbatively
finite parallel transport. This gives us a way to construct a theory
using using the state space of another a finite distance away. We
believe this result is of direct relevance to the difficulties found
in the analysis of background independence of [\rkz ] when discussing
string field theories formulated around conformal backgrounds a finite
distance apart. Moreover, having a perturbation construction manifestly
free of divergences, which was a main motivation in [\rsonoda ], should
be of utility for the problem of background independence where divergences
make the world sheet approach very difficult.

We would like to comment on the relation of the geometry
we have been discussing to the kind of geometry which has been explored
earlier. If we restrict ourselves to the tangent vector bundle of the
space of conformal field theories, we can construct
Riemannian geometry using the Zamolodchikov
metric [\rkutasov].  Related work for the case of $N=2$ theories and
a finite dimensional bundle has been discussed in [\rcecottivafa ].
In this paper we have studied geometry
of the {\it infinite} dimensional vector bundle which includes
the tangent vector bundle as a finite dimensional sub-bundle. (The
connection $\bar c$, restricted to the tangent bundle, gives the
usual Riemannian connection.)

We organize this paper as follows.  In \S2 we will summarize the
relevant features of the operator formalism for a space
of conformal field theories. In \S3 we establish that covariant
derivatives generate CFT deformations and explain Eqn.\xegen. In
\S4 we discuss some interesting connections that arise in a natural
way from the CFT data. This will lead us to introduce the connections
$\HG_D$, $c$, and $\bar c$. In \S5 we give algorithms for computing
the coefficients of connections. In \S6 we give our discussion and
computation of curvature, and in \S7 we study parallel transport
over finite distance. Finally, in \S8 we conclude the paper with
comments and open questions.

\chapter{Review and Notation}

\REF\ralvarez{L. Alvarez-Gaume, C. Gomez, G. Moore and C. Vafa, Nucl.
Phys. {\bf B303} (1988) 455.}

\section{Operator Formalism}

Let us recall the operator formalism for conformal field
theories.  (See, [\ralvarez ], and, sect.~2 of [\rcnw] for a summary.)
To define a conformal field theory, we introduce
an infinite dimensional linear space of ket-states, which
we denote by $\H$, and its dual space $\H^*$ of bra-states.
The same local field $\Phi_i$ corresponds to both the ket-state
$\ket{\Phi_i}$ and the bra-state $\bra{\Phi_i}$,
called the BPZ conjugate in [\zwiebachlong].
Let $\{ \ket{\Phi_i}\}$ be a basis of $\H$.
We introduce a dual basis $\{ \bra{\Phi^i} \}$ by
$$
\vev{\Phi^i \mid \Phi_j} = \delta_j^i .\eqn\edual
$$
In [\zwiebachlong], $\bra{\Phi^i}$ is called a conjugate state
of $\ket{\Phi_i}$.  Given a Riemann surface $\Sigma$
with punctures $P_1,...,P_n$ and local coordinates
$z_1,...,z_n$ such that $(\forall i) z_i(P_i) =0$,
the operator formalism assigns an element, $\bra{\Sigma;z_1,...,z_n}$,
in the $n$-th order tensor product
$\H^* \otimes ... \otimes \H^*$.  If $(\Sigma;z_1,...,z_n)$
and $(\Sigma';z_1',...,z_n')$
are analytically isomorphic as punctured Riemann surfaces with
coordinates, we find $\bra{\Sigma;z_1,...,z_n}
= \bra{\Sigma';z_1',...,z_n'}$.

One chooses a special two punctured sphere to define a metric.
Let $\Sigma$ be a two-punctured sphere with uniformizing coordinate
$z$, with a puncture at $z=0$ and a local coordinate $z_1=z$ at this
puncture, and a puncture at $z=\infty$ and a local coordinate
$z_2=1/z$ at this puncture. Let $\bra{R(z_1,z_2)}$ be the
state corresponding to this surface. We then define the metric
$$G_{ij} \equiv \bra{R(z_1,z_2)} \bigl( \ket{\Phi_i}
\otimes\ket{\Phi_j} \bigr) ,\eqn\defmet$$
where the state $\ket{\Phi_i}$ is inserted at $z_1=0$ and the state
$\ket{\Phi_j}$ is inserted at $z_2=0$.
This is usually called the Zamolodchikov metric, and it can be shown
to be symmetric using conformal invariance. The metric can be used
to lower the indices on the bra-states
$$ \bra{\Phi_i} \equiv \bra{\Phi^j} G_{ji}, \eqn\elowind$$
relating in this way conjugate states to the so-called BPZ conjugates.
It then follows that
$$G_{ij} = \bra{\Phi_i}\Phi_j\rangle .\eqn\ezmetric$$
A state related to the metric state above is the state $\ket{R}
\in \H \otimes \H$, defined by
$$\ket{R}\equiv\sum_{i,j}G^{ij}\ket{\Phi_i}\otimes\ket{\Phi_j} ,
\eqn\eR$$
where $G^{ij}$ is the inverse of $G_{ij}$. The states $\bra{R}$ and
$\ket{R}$ are BPZ conjugates of each other.

The other important state that we will often use is a standard
three punctured sphere. This state is denoted as $\bra{0, z', \infty^*}$ where
the puncture at $0$ has the standard coordinate $z$, the puncture at $\infty$
has the standard coordinate $1/z$, and the puncture at $z'$ has the
coordinate $z-z'$. The asterisk on $\infty$ denotes that for the
state space of that puncture we have turned the bra into a ket using
$\ket{R}$. Therefore to get a number we must insert a bra state at
infinity, as opposed to a ket state at zero. This three punctured sphere
will be used to define the operator product expansion in the next section.

The crucial property of the operator formalism is that
the states $\bra{\Sigma;z_1,...,z_n}$ are required to satisfy
the sewing rule: on a punctured surface $\Sigma_1 \infty
\Sigma_2$ obtained by identification $z_1 w_1 = 1$, we must find
$$
\bra{\Sigma_1 \infty \Sigma_2;z_2,...,z_{m},
w_2,...,w_n} = \bigl( \bra{\Sigma_1;z_1,...,z_m}\otimes
\bra{\Sigma_2;w_1,...,w_n}\bigr) \ket{R(z_1,w_1)} .\eqn\esewing
$$
Here $\ket{R(z_1,w_1)}$ converts a bra-state at $z_1 =0$ to
a ket-state at $w_1=0$, which is contracted with a bra-state at $w_1 =0$.
It has been shown that the sewing property is enough to
define a conformal field theory unambiguously.
(See, for example, [\rsegal].)
In the operator formalism
we obtain the correlator of local fields
$\Phi_{i_1}$, ..., $\Phi_{i_n}$ at points
$P_1$, ..., $P_n$ as an inner product
$$
\vev{\Phi_{i_1} (z_1=0) ... \Phi_{i_n} (z_n=0)}_\Sigma
= \bra{\Sigma;z_1,...,z_n} \left( \ket{\Phi_{i_1}}
\otimes ... \otimes \ket{\Phi_{i_n}} \right) .\eqn\ecorr
$$
In the rest of the paper we will denote $\bra{\Sigma;z_1,...,z_n}$
simply by $\bra{\Sigma}$ unless it is confusing to do so.

\subsection{Definition of the Virasoro operators}
In the operator formalism all that is given to us are the states
$\bra{\Sigma}$.  The Virasoro operators are not independent data
but can be constructed from these states. This was discussed
in [\rcnw]. We simply repeat from there the result for the
construction of $L_n$ from the two punctured sphere with punctures
at $0$ and $\infty^*$.
$$
L_n = {d \over d \epsilon_n} \bra{z + \sum_n \epsilon_n z^n ,\infty^*} \bigg|_{
\epsilon_n=0} \quad ,\eqn\edefln$$
where $z + \sum_n \epsilon_n z^n$ is the coordinate at zero, and
$1/z$ is the coordinate at infinity (recall that the asterisk denotes
that there we have a ket). Also
$$
\bar{L}_n = {d \over d \bar{\epsilon_n}} \bra{z + \sum_n
\epsilon_n z^n,\infty^*} \biggl|_{\bar \epsilon _n = 0} \quad .\eqn\edefbarln
$$
These expressions will be useful later when we want to compute the
covariant derivatives of the $L_n$.

\section{Operator Product Expansions}

In conformal field theory the operator product expansion
is always an exact statement. In the operator formalism
it arises because whenever an operator $\Phi_i(z)$ is inside the coordinate
disk $|z|\leq 1$ of another operator $\Phi_j(0)$ located at the origin,
their effect on the rest of the surface can be reproduced by sewing a
three punctured sphere onto the surface. This three punctured
sphere with uniformizing coordinate $w$ is punctured at $0,z$ and
$\infty$, and the local coordinates are $w,w-z$ and $1/w$ respectively.
The operator $\Phi_i$ is inserted at $w=z$, and $\Phi_j$ is inserted
at $w=0$. Such a three punctured sphere, since it has no insertion
at one puncture (the one at $w=\infty$), represents an element in
${\cal H}$, which is the element called the operator product expansion.
The constraints of dimensionality require that the operator product
expansion of a dimension $(1,1)$ field ${\cal O}_\mu (z,\bar z)$
(denoted for brevity as ${\cal O}_\mu (z)$) with an arbitrary
field $\Phi_i$ of dimension $(\Delta_i ,\overline \Delta_i )$
be of the form
$$
{\cal O}_\mu(z) \Phi_i (0) = {1\over 2\pi}\sum_k {H_{\mu i}^{~~k} \over
r^{2+\gamma_i-\gamma_k}}e^{-i\theta (s_i-s_k)} \Phi_k(0) ,\eqn\opeex
$$
where $\gamma_i = \Delta_i+ \overline\Delta_i$,
$s_i = \Delta_i -\overline\Delta_i$, and $z=r\exp (i\theta)$.
If the field ${\cal O}_\mu (z)$ is to be integrated over a region
with rotational invariance then we may simply use the expansion
$${\cal O}_\mu(z) \Phi_i (0) = {1\over 2\pi}\sum_k {H_{\mu i}^{~~k}
\delta_{s_i,s_k}\over
r^{2+\gamma_i-\gamma_k}} \Phi_k(0) .\eqn\opeex$$
A closely related expression -- the expression
for the three point function mentioned above -- is
$$\bra{\Phi^k}\,\bra{0,z,\infty^*}\ket{\Phi_i}\ket{\O_\mu}
= {1\over 2\pi} {H_{\mu i}^{~~k} \over |z|^{2+\gamma_i-\gamma_k}}
e^{-i\theta (s_i-s_k)}. \eqn\ethreept$$

We now define symbols $D_{\mu i}^{~~k}$ and
$F_{\mu i}^{~~k}$ by breaking up the sum over operators in the right
hand side of the operator product as follows
$${\cal O}_\mu(z) \Phi_i (0) = {1\over 2\pi}\sum_{\gamma_k\leq\gamma_i}
{D_{\mu i}^{~~k} e^{-i\theta (s_i-s_k)}\over
r^{2+\gamma_i-\gamma_k}}\, \Phi_k(0)
+{1\over 2\pi}\sum_{\gamma_k >\gamma_i}
{F_{\mu i}^{~~k} e^{-i\theta (s_i-s_k)}\over
r^{2+\gamma_i-\gamma_k}}\, \Phi_k(0) .\eqn\opeex$$
The notation $D$ and $F$ stand for divergent and finite, and
refer to the integrability of the singularity.
Note that each operator can only appear in one of the above sums.
We extend the definition of $D_{\mu i}^{~~k}$ and $F_{\mu i}^{~~k}$ by setting
them to zero for all $\gamma_k >\gamma_i$ and for all $\gamma_i\geq\gamma_k$
respectively
$$\eqalign{
D_{\mu i}^{~~k} &= 0, \quad\hbox{for}\,\gamma_k>\gamma_i ,\cr
F_{\mu i}^{~~k} &= 0, \quad\hbox{for}\,\gamma_k\leq\gamma_i .\cr}
\eqn\extdef$$
We consider $H_{\mu i}^{~~k}$ as a matrix with row index $k$
and column index $i$, and order the operators by increasing
dimensionality, that is $\gamma_i > \gamma_j$ when $i > j$.
Then we find that
$$H_{\mu i}^{~~k} = D_{\mu i}^{~~k} + F_{\mu i}^{~~k}\eqn\decope$$
is a decomposition of $H$ into an upper triangular matrix $D$ and a
matrix $F$ that has nonzero elements only below the diagonal.

\section{Geometry of Vector Bundles}

Let us consider a family of conformal field theories labelled by
continuous parameters $x^\mu (\mu = 1,...)$. One has a state space at
every point $x$, which constitutes a fiber of a vector bundle $V$
over theory space. As we have recalled in \S2.1, given an $n$
punctured Riemann surface $\Sigma$, with coordinates around
the punctures, a conformal field
theory at $x$ specifies a bra-state ${}_x\bra{\Sigma}$. This implies that we
have a section (of the $n$-tensor bundle) corresponding to every surface
$\Sigma$. Let {\it surface sections} denote such sections.

We denote local bases of $\H_x$ and its dual $\H^*_x$ by
$\{\ket{\Phi_i}_x \}$ and $\{{}_x\bra{\Phi^i}\}$.
We define a covariant derivative of an arbitrary ket-state $\ket{s(x)} \equiv
\sum_i s^i (x) \ket{\Phi_i}_x$,  as
$$D_\mu (\Gamma) \ket{s(x)} \equiv \sum_i
\biggl( \partial_\mu s^i (x)
+ \sum_j s^j (x) \Gamma_{\mu j}^{~~i} (x) \biggr)
\ket{\Phi_i}_x .\eqn\edmus$$
It will be convenient sometimes to use matrix notation. If we define
the connection as a matrix operator
$$\Gamma_\mu (x) = \Gamma_{\mu i}^{~~j}(x) \,
\ket{\Phi_j}\bra{\Phi^i},\eqn\mdef$$
then the covariant derivative of an arbitrary ket can be rewritten as
$$D_\mu (\Gamma) \ket{s(x)} =  \ket{\partial_\mu s (x)}
+ \Gamma_\mu (x)\ket{s(x)} .\eqn\xedmus$$
Under a change of the basis
$$\ket{\Phi_i}_x \to \sum_j N_i^{~j} (x) \ket{\Phi_j}_x ,
\eqn\echangeofbasis$$
the connection $\Gamma_\mu$ must transform as
$$\Gamma_\mu (x) \to N (x)
\left( - \partial_\mu + \Gamma_\mu (x) \right) N^{-1} (x)
\eqn\echangegamma$$
so that the covariant derivative in \edmus\ transforms
in the right way.
Likewise, we define a covariant derivative of
a bra-state by
$$
D_\mu (\Gamma) {}_x\bra{\Phi^i}
= - \sum_j {}_x\bra{\Phi^j} \Gamma_{\mu j}^{~~i} (x) .
\eqn\ecovbra
$$
Then, the covariant derivative of an arbitrary bra-state
$\bra{t(x)} \equiv \sum_i {}_x\bra{\Phi^i} t_i (x)$ is given by
$$D_\mu (\Gamma) \bra{t (x)} = \sum_i {}_x\bra{\Phi^i} \biggl( \partial_\mu t_i
(x) - \sum_j \Gamma_{\mu i}^{~~j} (x) t_j (x) \biggr) ,\eqn\edmut$$
and, in matrix notation
$$D_\mu (\Gamma) \bra{t(x)} = \bra{\partial_\mu t (x)}
 - \bra{t(x)}\Gamma_\mu (x) .\eqn\xedmut$$
A connection $\Gamma_\mu$ is called compatible with
the metric $G_{ij}$, if
$$
D_\mu (\Gamma) G_{ij} \equiv \partial_\mu G_{ij} -
\Gamma_{\mu i,j} - \Gamma_{\mu j,i} = 0, \eqn\ecompatibility$$
where $\Gamma_{\mu i,j} \equiv \sum_k \Gamma_{\mu i}^{~~k} G_{kj}$.
Finally, the curvature of a connection $\Gamma_\mu$ is defined by
$$\Omega_{\mu\nu i}^{~~~j} (\Gamma)
\equiv \partial_\mu \Gamma_{\nu i}^{~~j} - \partial_\nu \Gamma_{\mu i}^{~~j}
- \sum_k \bigl( \Gamma_{\mu i}^{~~k} \Gamma_{\nu k}^{~~j} -
\Gamma_{\nu i}^{~~k} \Gamma_{\mu k}^{~~j}\bigr) .\eqn\edefcurv$$
The curvature transforms covariantly under the change of basis
\echangeofbasis\
$$\Omega_{\mu\nu}(\Gamma)\to N\Omega_{\mu\nu}(\Gamma)N^{-1}.\eqn\etranscurv$$
As usual, curvature arises from the commutation of covariant
derivatives. A simple computation establishes that
$$[ D_\mu (\Gamma ) , D_\nu (\Gamma ) ]\, \bra{\Phi^i} \, =\,
-\bra{\Phi^k} \Omega_{\mu\nu k}^{~~~i} (\Gamma ).\eqn\curvaris$$
The generalization for an arbitrary (bra) tensor is simply
$$[ D_\mu (\Gamma ) , D_\nu (\Gamma ) ] \bra{s} =
-\bra{s} \sum_i \Omega_{\mu\nu}^{~(i)} (\Gamma ),\eqn\xcurvaris$$
where $\Omega_{\mu\nu}=\Omega_{\mu\nu i}^{~~j} \ket{\Phi_j}\cdot\bra{\Phi^i}$
is the curvature operator, and the superscript $(i)$ in the formula
labels the various state spaces of the tensor section.
This concludes our summary of some of the basic facts about
vector bundles. Most of our discusssions will be done in the language of
sections and will
therefore omit the label $x$. The presence of the label $x$ will indicate
that we are speaking of tensors and other objects in a particular state
space $\H_x$.

\section{Parallel Transport}

A large part of the physical discusson in the following sections
requires precise formulas for parallel transport. In this subsection
we will give the required results.

Let us consider a path $x^\mu (s)$ parametrized by $s\in [0,\e ]$.
Given a tensor $t(s=0)$ at $x^\mu (0)$, we want to define the tensor
$t(s)$ along the path by parallel transport. This is just the same
as requiring that the covariant derivative $D/Ds$ of $t(s)$ along the
path must vanish. For simplicity of notation assume $t$ is an arbitrary ket,
then the covariant derivative along the path is given by
$${D\over Ds} \ket{t(s)} \equiv {d \over ds}\ket{t(s)}
+ {dx^\mu \over ds}(s) \Gamma_\mu (s)\ket{t(s)}  = 0\, , \eqn\pathcov$$
where, as usual, the derivative $d/ds$ only differentiates the components
of the ket. Using component notation,
$\ket{t(s)} = t^i (s) \ket{\Phi_i}$, this equation reads
$$ {dt^i (s) \over ds} = -t^k (s) {dx^\mu \over ds}(s)
\Gamma_{\mu k}^{~~i} (s)\, ,\eqn\pathc$$
and in matrix notation we simply write
$$ {dt\over ds} = -t\, {dx^\mu \over ds}\, \Gamma_\mu \, .\eqn\xpathc$$
This, as usual, can be solved via a path ordered exponential
$$\eqalign{
t(s) \,\, &=\,\, t(0)\, {\cal P} \exp \left[
-\int_0^s ds'~ {d x^\mu (s') \over ds'}~
\Gamma_\mu (x(s')) \right]\cr
& =\,\, t(0) \Bigl( {\bf 1} - s {d x^\mu \over ds}(0) \Gamma_\mu (0)
- {s^2 \over 2} \Big[ {d^2 x^\mu \over ds^2}(0) \Gamma_\mu (0) \cr
&\quad\quad\quad + {d x^\mu \over ds}(0) {d x^\nu \over ds}(0) \bigl(
\partial_\mu \Gamma_\nu (0) - \Gamma_\mu (0)
\Gamma_\nu (0) \bigr) \Big] + ... \Bigr) ~.\cr}\eqn\eintegral$$
In the last step we have expanded the path ordered exponential
in terms of the standard nested integrals, and the first few integrals
were evaluated by expanding the integrand around the point $x(0)$
(with the path assumed to be real analytic). This formula
will be useful to analyze parallel transport beyond first order
in $s$.

Whenever we are given a vector {\it section} $t(x)$,
the covariant derivative $D/Ds$ along a path $x(s)$ is
simply given by
$${D t\over Ds} = {dx^\mu \over ds} \, D_\mu t ,\eqn\welll$$
and this formula is valid for any type of section (tensor type).
In fact, the notion of covariant derivative can be introduced
from that of parallel transport; for any section $t$, the
covariant derivative can be defined as
$$\delta x^\mu D_\mu\, t = t(x+\delta x \to x) - t(x) ,\eqn\ptlim$$
in the limit when $\delta x^\mu \to 0$. Here $t(x+\delta x\to x)$
is the value obtained when $t(x+\delta x)$ is transported to $x$
along the infinitesimal segment $\delta x$.
It follows from the above formula that
$$ t(x+\delta x \to x) =  t(x) + \delta x^\mu D_\mu\, t ,\eqn\xptlim$$
which simply says that one can do parallel
transport using a section and its covariant derivative.
Note that the left hand side is independent of the section used; it
only depends on the value of $t$ at the initial point and the connection
$\Gamma$. Therefore the right hand side, though not manifestly so,
is also independent of the values of $t$ in the neighborhood of
$x+\delta x$.

We denote the parallel transport
from $\H_{x+ \delta x}$ to $\H_x$ as the map $\T ( \Gamma)$. The above
equation implies that
$$\T(\Gamma ) : \,\,
\ket{\Phi_i}_{x+ \delta x}
\to \ket{\Phi_i}_x + \delta x^\mu D_\mu (\Gamma) \ket{\Phi_i}_x
= \ket{\Phi_i}_x + \delta x^\mu \Gamma_{\mu i}^{~~k} (x)
\ket{\Phi_k}_x .\eqn\emap
$$
For the dual basis vectors we
have a map from $\H^*_{x+\delta x}$ to $\H^*_{x}$ also denoted by $\T$
$$\T (\Gamma ) : \,\,
{}_{x+\delta x}\bra{\Phi^i}
\to {}_x\bra{\Phi^i} + \delta x^\mu D_\mu (\Gamma) {}_x\bra{\Phi^i}
= {}_x\bra{\Phi^i} - \delta x^\mu
{}_x\bra{\Phi^k} \Gamma_{\mu k}^{~~i} (x),\eqn\xemap$$
The map for the dual space was set up to preserve the contractions
between the state space and its dual
$\bra{\Phi^i}\Phi_j\rangle =\d^i_j$.
It follows that for an arbitrary vector section $\ket{t}$ and an
arbitrary dual vector section $\bra{s}$, the contraction, which
gives a number, is not altered by parallel transport. Indeed, under
such transport we have
$$\ket{t}_{x+ \delta x}
\to \ket{t}_x + \delta x^\mu D_\mu \ket{t}_x \,\, ; \quad\quad
{}_{x+\delta x}\bra{s}
\to {}_x\bra{s} + \delta x^\mu D_\mu \,\, {}_x\bra{s}\, ,\eqn\chcon$$
and therefore
$$\eqalign{
{}_{x+\delta x}\bra{s}t\rangle_{x+\delta x}\,\, \to \,\,
& {}_x\bra{s}t\rangle_x
+ \delta x^\mu \bigl(\, (D_\mu \bra{s} ) \ket{t}
+ \bra{s} (D_\mu\ket{t})\, \bigr) ,
\cr &= {}_x\bra{s}t\rangle_x + \delta x^\mu D_\mu \bigl(
{}_x\bra{s}t\rangle_x \bigr) ,\cr
&= {}_x\bra{s}t\rangle_x + \delta x^\mu \partial_\mu ({}_x\bra{s}t\rangle_x )
= {}_{x+\delta x}\bra{s}t\rangle_{x+\delta x} ,\cr}\eqn\nochange$$
as was expected. The first two lines of the above argument also
imply that the action of parallel transport {\it commutes} with
the operation of tensor contraction. If $s$ and $t$ are arbitrary
tensor sections, transporting and then contracting gives the same
result (tensor) as contracting and then transporting.

We now want to find the analog of \xptlim\ to all orders in $\delta x$.
Given an arbitrary section $t$ and a curve $x^\mu (s)$ we want
to parallel transport $t(x(0))$ along the curve. Let $\widetilde t (s)$
denote the tensor obtained by
paralell transport. It follows from \xptlim\ that
$$\widetilde t (\e ) = t (\e ) -\e {dx^\mu \over ds} (D_\mu t)(\e)
+ \O (\e^2) ,\eqn\fstep$$
where the covariant derivative in the right hand side has been evaluated
at $s=\e$ (which to $\O (\e^2 )$ is the same as evaluating it at $s=0$).
One can show, by iterating the infinitesimal transport equation, that the
finite version of the above is given by
$$\eqalign{\widetilde t(s) \, =\,  & t(s)
- \bigl( D_\mu t\bigr) (s) \int_0^s ds' {dx^\mu \over ds'} (s') \cr
& +\bigl( D_\mu D_\nu t\bigr) (s)
\int_0^s ds' {dx^\mu \over ds'} (s')
\int_0^{s'} ds'' {dx^\nu \over ds''} (s'')  +\cdots \cr
&+(-)^n\bigl( D_{\mu_1}\cdots D_{\mu_n} t\bigr) (s)
\int_0^s ds_1 {dx^{\mu_1} \over ds_1} (s_1)
\int_0^{s_1} ds_2 {dx^{\mu_2} \over ds_2} (s_2) \cdots
\int_0^{s_{n-1}} ds_n {dx^{\mu_n} \over ds_n}(s_n)+\cdots \,\cr}.\eqn\ptcovd$$
It should be noted that all the covariant derivatives are evaluated
at the final point of the path. It is straightforward to verify this
result by checking that ${D\widetilde t\over Ds} =0$. We can rewrite
this result in the language of path ordered exponentials
$$\widetilde t(s) \,\, =\, {\cal P} \exp \Bigl(
-\int_0^s ds'~ {d x^\mu (s') \over ds'}~
D_\mu  \Bigr) \,\, t(s) \, , \eqn\misled$$
where we keep in mind that the covariant derivatives act only
on the tensor $t(s)$.

\chapter{Connections and Deformations}

In \S2 we defined the vector bundle over theory space. Regarded as
a vector bundle it does not distinguish any connections, and one
must view all connections to be on an equal footing. However we have a vector
bundle with the CFT data -  a chosen family of sections $\bra{\Sigma}$. This
suggests that one might characterize covariant derivatives, $D_{\mu}(\Gamma)$,
in terms of their action on surface sections. For an arbitrary covariant
derivative $D_{\mu}(\Gamma)$, we will show that $D_{\mu}(\Gamma) \bra{\Sigma}$
generates a deformation of a conformal field theory.

In \S3.1 we will discuss and parametrize the relevant CFT deformations.
In \S3.2 we will prove that $D_{\mu}(\Gamma) \bra{\Sigma}$
generates a CFT deformation. This fact, combined with the parametrization
of CFT deformations, will lead us to the fundamental Eqn.\xegen .
In \S3.3 we will examine the ambiguities involved in associating with
a connection $\Gamma$ a pair $(D, \omega_\mu)$ that characterizes the CFT
deformations generated by $\Gamma$.

\section{CFT Deformations}

In this subsection we want to explain what a CFT
deformation is. Given a conformal theory a CFT deformation changes
(within the same state space) the states representing the surfaces
preserving the algebra of sewing. By definition, such deformations
give us a conformal theory if we start with one. There are two
types of deformations - those that do not change the underlying conformal
field theory and those that do. We will now explain this idea which will
lead us to a parametrization of the relevant CFT deformations.

The notion of a CFT deformation was clearly presented in [\rcnw].
The idea is to begin with a conformal field theory defined in some state
space. This means we have states
$\bra{\Sigma;z_1,...,z_m}$ satisfying the sewing property
$$
\bra{\Sigma_1\infty\Sigma_2} =
\bigl(\, \bra{\Sigma_1;z_1,...,z_m}\bigr) \bigl( \,\bra{\Sigma_2;w_1,...,w_n}
\, \bigr)
\ket{R(z_1,w_1)}.\eqn\esew
$$
We have a CFT deformation if we can define, for
every surface $\Sigma$, a deformed state $\bra{\Sigma,\epsilon}$,
in the same state space, so that the sewing property
is still satisfied. If
we write the deformed state as
$$
\bra{\Sigma,\epsilon} = \bra{\Sigma} + \epsilon \,\d \bra{\Sigma}\eqn\edef
$$
then the CFT condition
$$
\bra{\Sigma_1\infty\Sigma_2, \epsilon} =
\bigl(\,\bra{\Sigma_1;z_1,...,z_m, \epsilon}\bigr) \bigl(\,
\bra{\Sigma_2;w_1,...,w_n, \epsilon}\bigr)
\ket{R(z_1,w_1, \epsilon)}\eqn\edefsew
$$
translates into the requirement
$$
\eqalign {\d \bra{\Sigma_1\infty\Sigma_2} &=
(\d \bra{\Sigma_1;z_1,...,z_m}) \bra{\Sigma_2;w_1,...,w_n}
R(z_1,w_1)\rangle \cr
&\quad + \bra{\Sigma_1;z_1,...,z_m} (\d \bra{\Sigma_2;w_1,...,w_n})
\ket{R(z_1,w_1)} \cr
&\quad + \bra{\Sigma_1;z_1,...,z_m}\bra{\Sigma_2;w_1,...,w_n}
(\d \ket{R(z_1,w_1)}).\cr}\eqn\einfsew
$$
Equation \einfsew\ implies that if $\d_1 \bra{\Sigma}$ and $\d_2
\bra{\Sigma}$ are CFT deformations, then
$\alpha\, \d_1 \bra{\Sigma} + \beta \,\d_2
\bra{\Sigma}$ with $\alpha, \beta$, constants, is also a CFT
deformation .

It is important to recognize that in general the sewing state
$\ket{R(z_1,w_1)}$ changes under a deformation. In [\rcnw] attention was
restricted to deformations in which the sewing state remains the same.
There is no a priori reason to restrict our attention to such deformations. In
fact when we come to consider higher order deformations we will find it
essential to consider the more general deformations in
which $\ket{R(z_1,w_1)}$ changes.

Having defined a CFT deformation we note that all deformations may not change
the theory. An unchanged theory means that the
spectrum of operators is the same and the correlation functions are the same.
More precisely two theories $\bra{\Sigma,1}$ and $\bra{\Sigma,2}$ (described
in the same state space) are the same
if there is a one to one map (linear operator) of the state space onto
itself, such that the map it induces on tensors takes
$\bra{\Sigma,1}$ to $\bra{\Sigma,2}$ for every surface $\Sigma$. This
immediately suggests
a class of deformations that will not change the theory. Indeed choose an
arbitrary linear operator and let it act on the states
$\bra{\Sigma,1}$ as the generator of a linear map. From the above notion
of equivalence it follows that this deformation cannot change the theory.
Let us see explicitly.
An infinitesimal similarity transformation acts on the ket-states as
$$
\eqalign{
\ket{\Phi_i} \to
&\ket{\Phi_i} + \epsilon \omega \ket{\Phi_i} \cr
\quad \equiv & \ket{\Phi_i}
+ \epsilon \sum_j \omega_{i}^{~~j} \ket{\Phi_j},\cr}
\eqn\esimket
$$
while it acts on the dual bra-states as
$$
\eqalign{
\bra{\Phi^i} \to &
\bra{\Phi^i} - \epsilon \bra{\Phi^i} \omega \cr
\quad\equiv & \bra{\Phi^i}
- \epsilon\sum_j \bra{\Phi^j} \omega_{j}^{~~i}\cr}
\eqn\esimbra
$$
so that the duality $\vev{\Phi^i\mid\Phi_j} = \d_i^j$ is
preserved under the transformation. Eqn.\esimket\ implies that
the ket-state $\ket{R}$ transforms as
$$
\ket{R(z_1,z_2)} \to \ket{R(z_1,z_2)} + \epsilon \left( \omega^{(1)}
+ \omega^{(2)} \right) \ket{R(z_1,z_2)} \eqn\eesimR
$$
Similarly, \esimbra\ implies that the state $\bra{\Sigma}$, corresponding
to an $n$-punctured surface $\Sigma$, transforms as
$$
\bra{\Sigma} \to \bra{\Sigma} -
\sum_{i =1}^n \bra{\Sigma} \omega^{(i)} .\eqn\esimsigma
$$
{}From the fact that the similarity transformations cancel
out for the state spaces that are contracted,
we see that Eqn. \esimsigma\ is a CFT
deformation.  We have then a large class of CFT deformations that
do not change the theory.

Having identified all the deformations that do not change the CFT let us now
consider CFT deformations that do change the theory. A prescription for
these was given in Ref. [\rcnw], as
$$
\d^0 \bra{\Sigma} = - \hskip-6pt\int_{\Sigma -
\cup_i D_i}\hskip-6pt d^2 z~ \bra{\Sigma;z} \O \rangle , \eqn\ecnw
$$
where $\ket{\O}$ is the ket-state corresponding to a marginal
field $\O$ inserted at $z$, and the integral is performed over
$\Sigma$ with the unit disc around each puncture excluded. The requirement of
marginality is necessary if the integral on the right hand side of Eqn.\ecnw\
is to be well defined. Marginality makes the integrand independent of the local
coordinate used. Since the sewing of two surfaces is defined via
identification of two unit circles, the above rule manifestly yields a CFT
deformation. With this formula the sewing state $\ket{R(z_1,z_2)}$ does not
change. The reason for this is that the local coordinates cover the
whole surface and hence there is no area to integrate over.
Since deformations form a linear space we can modify \ecnw\  by the addition
of an arbitrary similarity transformation to get
$$
\d^{\omega} \bra{\Sigma} = \hskip-6pt\int_{\Sigma -
\cup_i D_i}\hskip-6pt d^2 z~ \bra{\Sigma;z} \O \rangle
- \sum_{i =1}^n \bra{\Sigma} \omega^{(i)} .\eqn\enewdef$$
Eqn.\enewdef\ parametrizes the space of CFT deformations.
Deformations are parametrized by choice of $\ket{\O}$ and
the pair $(D,\omega)$, where $D$ is the excluded domain and $\omega$
is an arbitrary linear operator.

\subsection {Ambiguities in the pair $(D, \omega)$}
There is overcounting in the parametrization of deformations since
a change in $D$ can be compensated by a change in $\omega$ to obtain
the same deformation. Let us see how this works.  For convenience
we take a domain $D$ and a domain $D'$ entirely included in $D$ ($D' \subset
D$).
The argument has a simple modifiction if $D'$ is not entirely included in $D$.
We wish to determine
$\omega' - \omega$ such that $(D, \omega)$ and $(D', \omega ')$ generate
the same deformation.
Let us start with the pair $(D, \omega)$ and break up the domain $D$ into
$D'$ and $D - D'$ to get
$$\eqalign{
 \delta^\omega \bra{\Sigma}
\, \equiv \, & - \hskip-6pt\int_{\Sigma -\cup D_i} \hskip-6pt d^2z
\bra{\Sigma ;z} \O (z)\rangle - \sum_i \bra{\Sigma} \omega^{(i)} \cr
= & -\int_{\Sigma -\cup D_i'} \hskip-6pt d^2z \bra{\Sigma ;z} \O (z)\rangle
+ \sum_i\int_{D_i - D_i'} \hskip-6pt d^2z \bra{\Sigma ;z} \O (z)\rangle
- \sum_i \bra{\Sigma} \omega^{(i)} \cr }
\eqn\rarss$$

Now in the integral over $D - D'$  replace the surface state
$\bra{\Sigma ;z}$ by the original state $\bra{\Sigma}$
with a standard three punctured sphere sewn to it at the puncture under
consideration to obtain
$$\bra{\Sigma ;z} = \bra{\Sigma} \bra{0,z,\infty^*},\eqn\sewin$$
where the asterisk denotes that this puncture is turned into a ket;
the ket is then
contracted with the state space in $\bra{\Sigma}$ that represented the original
puncture. It should be noted that exactly the same factorization holds
for every puncture. Therefore
$$\sum_i\hskip-4pt\int_{D_i-D_i '}\hskip-6pt d^2z \bra{\Sigma ;z}
\O (z)\rangle = \bra{\Sigma}\sum_i\hskip-3pt\int_{D_i-D_i '}
\hskip-6pt d^2z \bra{0,z,\infty^*} \O (z)\rangle ,\eqn\xrarss$$
where the integral in the right hand side is a matrix operator
in the state space $i$.  Substituting this in Eqn.\rarss we deduce
that $$\omega' - \omega = -\hskip-6pt\int_{D- D'} \hskip-5pt d^2z \,
\bra{0,z,\infty^*} \O (z)\rangle .\eqn\calcomega$$

There is a further ambiguity in specifying $\omega$ since one can
change it by the addition of a symmetry and the deformation will not
change. Let us recall how this works. We usually think of a
symmetry in a conformal theory as generated by dimension (1,0) or (0,1)
operators of theory. If $J_a(z)$ denotes a holomorphic
(or antiholomorphic) current then the Ward identity of this symmetry is
obtained by inserting the operator on a surface and integrating it around
each of the punctures. We then obtain the identity
$$
\bra{\Sigma,z_1 ... z_n} (J_{a,0} \ket{\Phi_1}) .... \ket{\Phi_n}+\,\,
\cdots\,\,+
\bra{\Sigma,z_1 ... z_n} \Phi_1\rangle .... (J_{a,0}\ket{\Phi_n})=0 ,
\eqn\esym$$
where $J_{a,0} = \oint {dz \over 2\pi i} J_a(z)$ is the zero mode of the
current. $J_{a,0}$ is a linear operator in the state space of the theory.
Since \esym\ holds for any set of states $\Phi_1 .... \Phi_n$ we can
rewrite \esym\ as
$$\sum_{i=1}^{n}\bra{\Sigma,z_1 ... z_n} J_{a,0}^{(i)} = 0. \eqn\esymgen$$
More generally, a symmetry is an operator which preserves correlation
functions when acting on the states of a theory. With this definition
in mind we see that any operator $S$ satisfying Eqn.\esymgen\
(with $S$ replacing $J_{a,0}$) is an infinitesimal
symmetry. The zero modes of (1,0) (or (0,1)) operators are
only special ways of constructing such symmetries. This concludes our
discussion of the ambiguities in specifying a pair $(D, \omega)$ corresponding
to a CFT deformation.

We have so far considered a fixed deformation and investigated different
ways of representing it in terms of $(D,\omega)$. This investigation allows
us to say more. If our interest is in the equivalence classes of deformations
- parametrized by $\ket{\O}$ and $(D,\omega)$ - that yield the same
deformed theory then we find that all deformations with the same
$\ket{\O}$ are in the same equivalence class. Indeed changing $\omega$
as stated earlier does not change the theory and we have just seen that
changing $D$ is equivalent to changing $\omega$. In other words the
equivalence classes of true deformations are parametrized
by the marginal operators - an expected conclusion.

\section{Covariant derivatives generate CFT Deformations}

We want to show now that an arbitrary covariant derivative
$D_\mu (\Gamma)$, generates a CFT deformation. In other words
we wish to show that whenever we deform states as
$$\bra{\Sigma,\epsilon} = \bra{\Sigma} + \epsilon \d x^\mu D_\mu (\Gamma)
\bra{\Sigma}\eqn\epartran$$
the sewing condition
$$
\bra{\Sigma_1\infty\Sigma_2, \epsilon} =
\bra{\Sigma_1;z_1,...,z_m, \epsilon} \bra{\Sigma_2;w_1,...,w_n, \epsilon}
R(z_1,w_1, \epsilon)\rangle ,\eqn\endefsew$$
is satisfied. The proof is essentially a triviality once we realize
what \epartran\ means geometrically. It follows from \xptlim\ that the
right hand side of Eqn.\epartran\ is simply the surface state
obtained by parallel transporting, with $\Gamma$, the state $\bra{\Sigma}$
at $x+\e\d x$, to $x$ (to first order in $\d x$).
Therefore, our explanations below Eqn.\nochange\ suffice; sewing
corresponds to tensor contraction, and our deformation above is
generated by paralel transport; since parallel transport and
tensor contraction commute, the deformations generated by covariant
derivatives are CFT deformations for any choice of connection.

We must ask what $\ket{\O}$ and $(D,\omega)$ correspond
to these deformations. The existence of a family of
CFT's implies a map from the tangent space at $x$ to
the marginal states $\ket{\O}$ in $\H_x$. We therefore can speak
of the marginal states $\ket{\O_\mu}$ corresponding to the basis
vectors in the tangent
space. In fact, we take as the definition of a family of conformal theories
the statement that for any connnection $\Gamma$,
$D_\mu (\Gamma) \bra{\Sigma}$ is generated by $\ket{\O_\mu}$. This means
$$D_\mu (\Gamma) \bra{\Sigma} = - \hskip-6pt \int_{\Sigma -
\cup_i D_i}\hskip-6pt d^2 z~ \bra{\Sigma;z} \O_\mu \rangle
-  \sum_{i=1}^n \bra{\Sigma}\omega_\mu^{(i)},\eqn\egen$$
for some choice of an operator one form $\omega_\mu$ and domain
$D$ at each point in theory space. This is an important equation. It
will be the starting point of much of our later work. It explains the relation
between covariant derivatives and deformations, and, as we will see, the
variational formula postulated in  ref.~[\rsonoda] follows directly from it.
It is appropriate to regard this formula as part of the very definition of
a space of conformal theories.

We note that the space spanned by the $\ket{\O_\mu}$'s might only be a subspace
of the space of marginals. This idea has been elaborated in refs. [\rdijk].
The $\ket{\O_\mu}$'s correspond to the exactly marginal
states. We will recover later in our analysis some of the conditions
that must be satisfied by the exact marginals.

Let us see how equation \egen\ works as we change the connection $\Gamma$. For
any connection $\Gamma '=\Gamma + \Delta\Gamma$  we have by definition that
$$D_\mu (\Gamma + \Delta\Gamma )\bra{\Sigma} = D_\mu (\Gamma ) \bra{\Sigma}
- \sum_{i=1}^n \bra{\Sigma} {\Delta\Gamma}^{(i)}.\eqn\tradeoff$$
If Eqn. \egen\ holds for the connection $\Gamma$ then Eqn. \tradeoff\ implies
that for the connection $\Gamma '$ we find
$$
D_\mu (\Gamma ') \bra{\Sigma} =  -  \hskip-6pt
\int_{\Sigma -\cup_i D_i}\hskip-6pt d^2 z~ \bra{\Sigma;z}
\O_\mu \rangle - \sum_{i =1}^n
\bra{\Sigma} \omega '^{(i)}_\mu ,\eqn\echgamma$$
with $\omega_\mu ' =  \omega_\mu + (\Gamma ' - \Gamma)$.

\subsection{Ambiguities in the pair $(D, \omega_\mu)$}
As in the case of deformations ( Eqn.\enewdef), Eqn.\egen\ also implies an
ambiguity in the association of a covariant derivative with a pair
$(D, \omega_\mu)$. The only difference is that $D$ and
$\omega_\mu$ can be chosen at each point in theory space.
Again, a change in $D$ can
be compensated by a change in $\omega_\mu$. It is also clear that $(D,
\omega_\mu + s_\mu)$
is another pair associated with $D_\mu(\Gamma)$ if $(D, \omega_\mu)$ is.
Here $s_{\mu}$ is a symmetry one form, that is, an operator
for which $s_{\mu i}^{~~j}\, \d x^\mu$ is a symmetry for any choice
of tangent vector $\d x^\mu$.
Besides the ambiguity in associating a pair $(D, \omega_\mu)$ with a connection
$\Gamma$ there is an ambiguity in the reverse direction as well due to the
presence of symmetries. From \echgamma\ we see that
$$
D_\mu (\Gamma + s) \bra{\Sigma} =  D_\mu (\Gamma) \bra{\Sigma}, \eqn\sym
$$
This implies that given a pair $(D, \omega_\mu)$, the corresponding connection
coefficients $\Gamma_{\mu i}^{~~j}$ are ambiguous up to the addition of
symmetry coefficients $s_{\mu i}^{~~j}$. This fact will be relevant
in \S5.

\chapter{Canonical Deformations and Connections}

In the previous section we have found that covariant derivatives generate
CFT deformations and that allowed us to identify clearly the role of
connections.
We will now investigate whether any covariant derivatives
are distinguished by the CFT data. We have approached
the issue of distinguishing connections in the spirit of exploring
the possible connections and pointing out some that have interesting
characteristics.
Without the CFT data we characterize connections in
terms of metric compatibility, curvature, and finiteness of parallel
transport; the presence of CFT data gives a new characteristic - the
pair $(D, \omega_\mu)$ that determines the covariant derivative of
surface sections. It is this additional characteristic that will be
the focus of our attention. There are pairs $(D, \omega_\mu)$ that are
natural in the manner of their construction. The natural connections, then,
are those that are associated with these pairs.

\section{The connections $\HG_D$}

We see that specifying the right hand side of Eqn.\egen\ amounts
to choosing a pair $(D, \omega_\mu)$. In general we need to make choices for
this pair at each point $x$ in theory space. There are some deformations
that can be regarded as constant. An analogy is in order. If we
are asked to choose functions on a manifold we have choices to make at each
point on the manifold. However the constant functions are special in
that there are fewer choices to make. In the same way if we fix the
domain to be the same at all points and set $\omega_\mu = 0$,
then we have obtained the analog of the constant functions. The connections
that yield such deformations will be denoted $\HG_D$
 $$D_\mu (\HG_D) \bra{\Sigma} =
-\hskip-6pt\int_{\Sigma - \cup_i D_i}d^2z \, \bra{\Sigma;z}
{\cal O}_\mu (z)\rangle . \eqn\xwarf$$
One particular case of interest is the case where $D = D^1$ (the unit disc).
The corresponding connections will be denoted $\HG$.

We will now show, on general grounds, that $\HG$ is always compatible
with the metric $G_{ij}$.
If we consider a two-punctured sphere $\Sigma$, with uniformizing coordinate
$z$ and local coordinates $z_1 = z$ and $z_2 = 1/z$, around $z=0$ and
$z=\infty$ respectively, then the unit disks around the punctures
cover precisely the sphere. This implies that $D_\mu (\HG) \bra{\Sigma}=0$.
Given that the bra $\bra{\Sigma}$ can be written as
$\bra{\Sigma} = G_{ij}\bra{\Phi^i}\bra{\Phi^j}$, the vanishing of its
covariant derivative is, by definition, the statement of metric
compatibility.

\section {The connections $c$ and $\bar c$}

When a space of conformal theories is constructed from path integrals
we expect that the derivative of a correlation function will be
given by an integral of the conjugate operator $\O_\mu$ over the
entire Riemann surface
without any excluded domain. This is too naive, however, due to the
presence of divergences when the operator approaches another one. A
way that this might be corrected is to subtract just the divergent parts
away. This was the motivation that led [\rsonoda] to the introduction of
the variational formula with the connection $c$.
In this subsection we will switch from the operator
formalism to the language of correlation functions. Though just a
change in notation, it is useful to be able to move from one to
the other. This was the language used in
refs.~[\rsonoda].

We consider a set of fields
$\Phi_i(x)$ with $i\in S$ (a set) inserted on the surface $\Sigma$
with local coordinates $z_i$. If we expand a surface state, say, with
$n$ punctures as
$$\bra{\Sigma} = \sum_{i_1\cdots i_n} \bra{\Phi^{i_1}(x)}\otimes \cdots
\otimes \bra{\Phi^{i_n}(x)}\,\Sigma_{i_1\cdots i_n}(x) ,\eqn\exssta$$
the coefficients $\Sigma_{i_1\cdots i_n}$ of the expansion
are given by the correlators of local fields
$$\Sigma_{i_1\cdots i_n}(x) = \bra{\Sigma}\Bigl( \ket{\Phi_{i_1}(x)}
\otimes\cdots\otimes \ket{\Phi_{i_n}(x)}\Bigr) \equiv
\bigl\langle \Phi_{i_1} \cdots \Phi_{i_n} \bigr\rangle_\Sigma .\eqn\decomm$$
Conventionally, we define
$$D_\mu \Sigma_{i_1\cdots i_n} \equiv \bigl( D_\mu \bra{\Sigma} \bigl)
\ket{\Phi_{i_1}}\otimes\cdots\otimes\ket{\Phi_{i_n}}. \eqn\cnvdef$$
This gives, as usual,
$$D_\mu \Sigma_{i_1\cdots i_n}=
\partial_\mu \Sigma_{i_1\cdots i_n} - \Gamma_{\mu i_1}^{~~k}
\Sigma_{k\cdots i_n} -\cdots - \Gamma_{\mu i_n}^{~~k}
\Sigma_{i_1\cdots k}.\eqn\asusual$$
Therefore, in the language of correlators, our equation \egen\ reads
$$D_\mu(\Gamma ) \,\bigl\langle \prod\limits_{i\in S} \Phi_i (z_i)
\bigr\rangle_{\Sigma} =
-\hskip-8pt\int_{\Sigma - \cup_i D_i}\hskip-8pt d^2z \,\bigl\langle {\cal
O}_\mu (z)
\prod\limits_{i\in S} \Phi_i (z_i) \bigr\rangle_{\Sigma} - \sum_{i\in S}
\sum_k \omega_{\mu i}^{~~k} \cdot \bigl\langle \Phi_k(z_i) \prod\limits_{{j\in
S}
\atop {j\not= i}}\Phi_j(z_j) \bigr\rangle_\Sigma . \eqn\ppcvarf$$
Expanding out, this formula reads
$$\partial_\mu \,\bigl\langle \prod\limits_{i\in S} \Phi_i (z_i)
\bigr\rangle_\Sigma =
-\hskip-8pt\int_{\Sigma - \cup_i D_i}
\hskip-8pt d^2z \,\bigl\langle {\cal O}_\mu (z)
\prod\limits_{i\in S} \Phi_i (z_i) \bigr\rangle_\Sigma
+ \sum_{i\in S} \sum_k \left(\Gamma_{\mu i}^{~~k} - \omega_{\mu i}^{~~k}\right)
\cdot \bigl\langle \Phi_k(z_i) \prod\limits_{{j\in S}\atop {j\not= i}}
\Phi_j(z_j) \bigr\rangle_\Sigma .\eqn\pcvarf$$

\subsection{The connection $c$.}
The connection $c$ [\rsonoda ] is defined with an
excluded domain $D$ that tends to zero, and an $\omega_\mu$ that corresponds
to subtracting away the divergences as $\O_\mu(z)$ approaches each of the
punctures. Whenever $\O_\mu$ approaches a puncture we can use the OPE
expansion to write the subtraction term. We take
$$\eqalign{
D_\mu (c) \,\bigl\langle \prod\limits_{i\in S} \Phi_i (z_i)
\bigr\rangle_\Sigma = &\lim_{\epsilon \rightarrow 0} \, \biggl[ \,
-\int_{\Sigma - \cup_i D_i^\epsilon}
\hskip-6pt d^2z \,\bigl\langle {\cal O}_\mu (z)
\prod\limits_{i\in S} \Phi_i (z_i) \bigr\rangle_\Sigma \cr
&+ \sum_{i\in S} \sum_k
\hskip-4pt \int_{D_i-D_i^\epsilon}\hskip-10pt
{d^2z\over 2\pi}{D_{\mu i}^{~~k}\delta_{s_k,s_i}\over r^{2+\gamma_i-\gamma_k}}
\cdot \bigl\langle
\Phi_k(z_i) \prod\limits_{{j\in S}\atop {j\not= i}}\Phi_j(z_j)
\bigr\rangle_\Sigma \,\biggr] \, .\cr} \eqn\ppxtvarf$$
The choice of excluded domain and subtraction can be read by comparison
with Eqn.\ppcvarf . Another way to think of this definition is that we always
integrate the insertion over $\Sigma - \cup_i D_i^1$, that is over the surface
minus the unit disks. In addition, we do selective integration over the
disks. Whenever ${\cal O}_\mu (z)$ enters a disk $D_i^1$ we use the
operator product expansion of $\O_\mu$ with the operator sitting
at the puncture, thus inducing a sum of operators.
For operators appearing with nonintegrable singularities we simply
do not integrate any further, for operators appearing with integrable
singularities we integrate over all the disk.

Since for nonintegrable singularities we do not integrate beyond the
unit disc, $c$ is related to the connection $\HG$ (where no operator
is integrated beyond the unit disks) by the integral of the finite part
of the operator product expansion, that is,
$$c_{\mu i}^{~~k} = \HG_{\mu i}^{~~k}
+ \int_{D^1_i} {d^2z\over 2\pi}
{F_{\mu i}^{~~k} \delta_{s_i,s_k}\over r^{2+\gamma_i-\gamma_k}} .\eqn\conson$$
For $\gamma_i \geq \gamma_k$, $F_{\mu i}^{~~k}=0$ and we have
$$c_{\mu i}^{~~k} = \HG_{\mu i}^{~~k}\quad \hbox{for}\,\gamma_i\geq\gamma_k.
\eqn\relconni$$
For $\gamma_i < \gamma_k$ we have
$$\int_{r\leq 1}{d^2z\over 2\pi}{F_{\mu i}^{~~k} \delta_{s_i,s_k}
\over r^{2+\gamma_i-\gamma_k}}
=\int{d\theta\over 2\pi} \int_0^1 dr
{H_{\mu i}^{~~k} \delta_{s_i,s_k}
\over r^{1+\gamma_i-\gamma_k}}= - {H_{\mu i}^{~~k}\delta_{s_i,s_k} \over
\gamma_i - \gamma_k},\eqn\simpl$$
where Eqn.\opeex\ was used. We will show in the next section
(Eqn. (5.10)) that
$$
\HG_{\mu i}^{~~k} = {H_{\mu i}^{~~k}\delta_{s_i,s_k}
 \over \gamma_i - \gamma_k}\, ,\quad\hbox{for}\quad
\gamma_i\not= \gamma_k .\eqn\trmofconn
$$
Substituting Eqn.\trmofconn\ into Eqn.\conson\ we obtain
$$c_{\mu i}^{~~k} = 0 , \quad \hbox{for}\, \gamma_i <\gamma_k .\eqn\rest$$
Thus the connection $c_\mu$ is upper triangular.

\subsection{The connection $\bar c$.} There is yet another interesting
connection where we still integrate the
insertion all over the surface, but the subtractions $\omega_\mu$
are modified. We saw that the $\omega_\mu$ for the connection $c$ was
given in terms of an integral involving the upper triangular matrix $D_\mu$.
The integral can be performed to obtain
$$\eqalign{
\sum_k \hskip-4pt \int_{D_i-D_i^\epsilon}\hskip-10pt {d^2z\over 2\pi}
{D_{\mu i}^{~~k}\delta_{s_k,s_i}\over r^{2+\gamma_i-\gamma_k}}
&=\biggl( \sum_{\gamma_k < \gamma_i} +
\sum_{\gamma_k = \gamma_i}\biggr) \int_\epsilon^1 dr
{D_{\mu i}^{~~k}\delta_{s_k,s_i}\over r^{1+\gamma_i-\gamma_k}}\cr
&=-\sum_{\gamma_k < \gamma_i}
{D_{\mu i}^{~~k}\delta_{s_k,s_i}\over \gamma_i-\gamma_k}
\Bigl(1 - {1\over \epsilon^{\gamma_i- \gamma_k}} \Bigr)
-\sum_{\gamma_k = \gamma_i}
D_{\mu i}^{~~k}\delta_{s_k,s_i}\ln \epsilon .\cr}\eqn\diagcon$$
We recognize that the first term in the last right hand side
has a finite part in $\epsilon$ which is nothing else than
the above-diagonal part of the upper triangular connection $c$.
We may do a `minimal subtraction' where we only  subtract
the divergences in the expansion in $\epsilon$ and retain the
finite terms. This suggests that we define a connection $\bar c$
which satisfies
$$\eqalign{
D_\mu (\bar c) \,\bigl\langle \prod\limits_{i\in S} \Phi_i (z_i)
\bigr\rangle_\Sigma &= \lim_{\epsilon \rightarrow 0} \, \biggl[ \,
-\int_{\Sigma - \cup_i D_i^\epsilon}
\hskip-6pt d^2z \,\bigl\langle {\cal O}_\mu (z)
\prod\limits_{i\in S} \Phi_i (z_i) \bigr\rangle_\Sigma \cr
&\hskip-15pt + \sum_{i\in S}
\biggl(  \sum_{\gamma_k < \gamma_i}
{D_{\mu i}^{~~k}\delta_{s_k,s_i}\over \gamma_i-\gamma_k}
{1\over \epsilon^{\gamma_i- \gamma_k}}
-\sum_{\gamma_k = \gamma_i}
D_{\mu i}^{~~k}\delta_{s_k,s_i}\ln \epsilon  \biggr)
\cdot \bigl\langle
\Phi_k(z_i) \prod\limits_{{j\in S}\atop {j\not= i}}\Phi_j(z_j)
\bigr\rangle_\Sigma \,\biggr]. \cr} \eqn\ppxyy$$
Such a `minimal' connection $\bar c$ is diagonal and equals the
diagonal part of the connection $\HG$
$$\bar c_{\mu i}^{~~k} \equiv
c_{\mu i}^{~~k} \,\delta_{\gamma_i ,\gamma_k}
= \HG_{\mu i}^{~~k}\delta_{\gamma_i ,\gamma_k}, \eqn\dvcx$$
The diagonal connection $\bar c$ is compatible
with the metric. Indeed
$$ D_\mu (\bar c) G_{ij} \equiv \partial_\mu G_{ij}
-{\bar c}_{\mu i}^{~~k} G_{kj} - {\bar c}_{\mu j}^{~~k} G_{ik} = 0 ,
\eqn\diagconnmc$$
because when $\gamma_i\not= \gamma_j$ all three terms vanish given
that both $G$ and $\bar c$ are diagonal, and, if $\gamma_i = \gamma_j$,
the vanishing follows because for all diagonal elements
$\bar c = \HG$, and $\HG$ was already shown to be metric
compatible.

We will write  Eqn.\ppxyy\  in the operator
formalism since it will be of use later. We first rewrite
it as
$$\eqalign{
D_\mu (\bar c) \,\bigl\langle \prod\limits_{i\in S} \Phi_i (z_i)
\bigr\rangle_\Sigma &= \lim_{\e \to 0} \, \biggl[ \,
-\int_{\Sigma - \cup_i D_i^\epsilon}
\hskip-6pt d^2z \,\bigl\langle {\cal O}_\mu (z)
\prod\limits_{i\in S} \Phi_i (z_i) \bigr\rangle_\Sigma \cr
&\hskip-5pt + \sum_{i\in S} \D_{\mu i}^{~~k} (\epsilon )
\cdot \bigl\langle
\Phi_k(z_i) \prod\limits_{{j\in S}\atop {j\not= i}}\Phi_j(z_j)
\bigr\rangle_\Sigma \,\biggr] \cr} ,\eqn\ppxyy$$
where we have defined
$$\D_{\mu i}^{~~k} (\epsilon ) =
\cases{ 0 ,&if $\gamma_k >\gamma_i$; \cr
{ H_{\mu i}^{~~k}\delta_{s_k,s_i} \over \gamma_i-\gamma_k }
{ 1\over \epsilon^{\gamma_i- \gamma_k} },
&if $\gamma_k <\gamma_i$; \cr
H_{\mu i}^{~~k}\delta_{s_k,s_i}\ln \epsilon ,&if
$\gamma_k=\gamma_i$ .\cr}\eqn\expsub$$
We then define the operator
$$\D_\mu (\e ) =  \ket{\Phi_k} \cdot \bra{\Phi^i} \, \D_{\mu i}^{~~k} (\e ),
\eqn\opedd$$
and can now write Eqn. \ppxyy\ as
$$ D_\mu (\bar c ) \bra{\Sigma} =
\lim_{\e \to 0} \, \bigl[ \,
-\hskip-8pt\int_{\Sigma - \cup_i D_i^\epsilon}
\hskip-8pt d^2z \, \bra{\Sigma ;z} \O_\mu (z)\rangle
+ \bra{\Sigma} \, \sum_i \D_\mu^{(i)} (\e ) \bigr] ,\eqn\operder$$
where the $i$ index in the second term refers to the puncture
(or state space) where the operator is inserted.

This completes the discussion of the special connections, the most
interesting of which will be the connections $\HG , c$ and $\bar c$.
We may regard the three of them as connections associated with a pair
$(D, \omega_\mu)$ where $D$ is zero and $\omega_\mu$ is the subtraction
of three point functions integrated over the unit disc. For $\HG$
we subtract all matrix elements, for $c$
we subtract those matrix elements that have divergences, and for $\bar c$
we subtract only the divergent part of the divergent matrix elements.

\chapter{The Connection Arising from a Deformation}

We have so far taken an arbitrary covariant derivative and considered its
action on the surface sections $\bra{\Sigma}$. We saw that this allows us
to associate with the covariant derivative the pair $(D, \omega_\mu)$.
We can now invert the process. We
consider a particular pair $(D, \omega_\mu)$ and ask for a
covariant derivative $D_\mu (\Gamma)$ associated
to that pair. For example, we may ask for the explicit form of the
connection $\HG$ which is associated with the pair $(D^1, 0)$.

This problem of inversion is the subject of this section.
We begin with some observations about the nature
of the problem which will allow us to formulate the problem precisely.
We will proceed through the use of the Virasoro operators.
In particular we will compute the covariant derivative of the Virasoro
operators.
This information will be used to determine the connection coefficients
$\Gamma_{\mu i}^{~~j}$ for $\gamma_i \neq \gamma_j$ in terms of OPE
coefficients and $\omega_\mu$. All other connection coefficients
follow from the knowledge of the connection
coefficients coupling primary states of the same dimensionality.

\section{Formulation of the Problem}

Since two connections differing by a symmetry generate the same
deformation, a deformation cannot fix the connection completely
unless there are no symmetries.  Once a basis is chosen,
if the matrix element of $s_{\mu i}^{~~j}$ between two basis states
$\Phi_i$ and $\Phi_j$ is zero then $\HG_{\mu i}^{~~j}$ is unambiguous.
Hence all such matrix elements can be determined in principle.

Our problem is that of a set of linear equations for
$\HG_{\mu i}^{~~j}$, one for every choice of surface with local coordinates
around the punctures.
Hence we have an overdetermined system of equations. However there
is no problem regarding the existence of a solution to this system
of equations due to the argument leading to Eqn.\egen .
Given an arbitrary $\Gamma$ there is an $\omega$ such that \egen\ holds.
Since any two CFT deformations (associated to $\delta x^\mu$)
can {\it only} differ by a choice of $\omega$, and a change of $\omega$ can be
traded by a change in $\Gamma$, there is some connection that yields any
CFT deformation we may choose.
It must be emphasized that while our concrete discussion in the present
section will be carried out for the pair $(D^1, 0)$ giving rise
to the covariant derivative $D_\mu(\HG )$, the strategy applies to any
connection.

There are then two steps
involved in obtaining information from this system of equations.
First we must choose a basis. This determines which matrix elements
of $\HG$ can even in principle be determined. Second, we must
choose a suitable set of equations so that we can solve for some
or all of these solvable coefficients. Our goal is then to choose a
basis with the greatest number of solvable coefficients and choose
equations that are the simplest to solve.

\section{Using the Virasoro operators}

One way to make the two choices listed above is through use of the
Virasoro operators. The motivation is as follows. If we choose a Lie
algebra of operators which commute with all the symmetries, then the
symmetries will not  mix representations that are inequivalent and the
mixing within the same kinds of representation happens in a manner that is
constrained and known. One can then choose a basis within each irreducible
representation which when combined will yield an overall
basis in which we can easily identify some matrix elements of
$\HG_{\mu i}^j$ that can in principle be determined. In using this observation
we will be assuming that the same Lie algebra is represented in each of the
representation spaces.

\subsection{Choice of basis}
An obvious choice for such a Lie algebra is the Virasoro algebra which always
commutes with the symmetries
$$[\, S, L_n\, ] = 0.\eqn\symcomm$$
To see why this is true we note that the $L_n$'s (as reviewed in
\S2, Eqns. \edefln ,\edefbarln) are defined intrinsically via two punctured
spheres with one puncture representing a bra and the other a ket. Thus
invariance under symmetry of the surfaces implies that
$$S\, L_n\, S^{-1} = L_n \, , \eqn\actsthis$$
since if we use $S$ to transform the kets, we must use $S^{-1}$
to transform the bra. Equation \symcomm\ follows immediately from \actsthis .

Having seen that the $L_n$'s commute with any symmetry $S$ we see that
choosing a basis constructed from irreducible representations of the Virasoro
algebra will be an option independent of the actual symmetries of the theory.
An
arbitrary element of the basis is labelled as
$$
L_{-n_1}L_{-n_2}\cdots L_{-n_j}\ket{h,\bar{h},i}
$$
where $\ket{h,\bar{h},i}$ is a primary state of dimension $(h,\bar{h})$ and $i
= 1 ... m$ indexes the the $m$ different primaries of this dimension.

\subsection{Choice of Equations.}
Having chosen a basis let us now  address the choice of equations for the
computation of the matrix elements of the connection.
The equations we will choose are obtained by taking the covariant derivatives
of
the sections corresponding to the Virasoro operators. When studying the
resulting equations we will write them out with a basis determined by the
Virasoro operators, as discussed earlier.

We do our computation for the connection $\HG$. Since this connection is
associated to the pair $(D^1, 0)$ we can obtain the covariant derivative
of the $L_n's$ from Ref.[\rcnw] (Eqn.(3.1.4)).  Using our normalization,
the result is
$$
D_\mu (\HG)\, L_n
= \oint_{|z|=1} {d \bar{z} \over - 2 i}~ z^{n+1}\, \bra{0,z, \infty^*}
\O_\mu\rangle ,
\eqn\edeltaLnformula
$$
where we orient the contour by $\oint d\bar{z}/\bar{z} = - 2 \pi i$.
In components this equation reads
$$\eqalign{
\left( D_\mu (\HG) L_n \right)_i^{~j}
& = \oint_{|z|=1} {d \bar{z} \over - 2 i}~ z^{n+1} \vev{
\O_\mu (z) \Phi_i (0) \Phi^j (\infty)} \cr
& = \int_0^{2 \pi} {d\theta \over 2}~{\rm e}^{i n \theta}\sum_k {H_{\mu
i}^{~~j}
\over 2 \pi}~{\rm e}^{ - i ~(s_i - s_j) \theta}
= {1 \over 2}~H_{\mu i}^{~~j} \d_{s_i,s_j+n} ,\cr}\eqn\eniceformula$$
where we have used the OPE \opeex.
Similarly, by choosing $\bar{v}$ non-vanishing instead
of $v$, we obtain another Virasoro condition. All in all we have
$$\eqalign{
\left( D_\mu (\HG) \overline{L}_n \right)_i^{~j} &= {1 \over 2} ~
H_{\mu i}^{~~j} \,\d_{s_i,s_j-n},\cr
\left( D_\mu (\HG) L_n \right)_i^{~j} &= {1 \over 2} ~
H_{\mu i}^{~~j}\, \d_{s_i,s_j+n}.\cr}\eqn\ealsoniceformula$$
These are the equations expressing the covariant derivatives
of the Virasoro operators.

Expanding out the expressions for the covariant derivatives as
$$
\left( D_\mu (\HG) L_n \right)_i^{~j}
= \partial_\mu \left( L_n \right)_i^{~j} -
\HG_{\mu i}^{~~k} \left( L_n \right)_k^{~j}
+ \left( L_n \right)_i^{~k} \HG_{\mu k}^{~~j} ,\eqn\whynoname$$
substituting into Eqn.\ealsoniceformula\  we find
$$\left( L_n \right)_i^{~k}\, \HG_{\mu k}^{~~j}
-\HG_{\mu i}^{~~k}\, \left( L_n \right)_k^{~j}
= [ \, L_n\, , \,\HG_\mu \, ]_i^{~j} = \partial_\mu \left(
L_n \right)_i^{~j} - {1 \over 2} ~
H_{\mu i}^{~~j}\, \d_{s_i,s_j-n}.\eqn\mustana$$
This equation and its analog can be studied to obtain some of the
coefficients. We write them as follows
$$\eqalign{
\left( L_n \right)_i^{~k}\, \HG_{\mu k}^{~~j}
+ \bigl( \delta_i^k \,\partial_\mu  - \HG_{\mu i}^{~~k} \bigr) (L_n)_k^{~j}
&= {1\over 2} H_{\mu i}^{~j}\, \delta_{s_i,s_j+n} \cr
\left( \overline L_n \right)_i^{~k}\, \HG_{\mu k}^{~~j}
+ \bigl( \delta_i^k\, \partial_\mu  - \HG_{\mu i}^{~~k} \bigr)
(\overline L_n)_k^{~j}
&= {1\over 2} H_{\mu i}^{~j}\,\delta_{s_i,s_j-n}. \cr}\eqn\freeal$$

\subsection{Solving the Equations}

We now have the equations that we wish to solve. When $n=0$ the above
Virasoro conditions give us the following equations
$$
\eqalign{\gamma_i \,\HG_{\mu i}^{~~k} + (\d_i^k\, \partial_\mu  - \HG_{\mu
i}^{~~k})\gamma_k &= H_{\mu i}^{~~k}\, \d_{s_k,s_i}\cr
s_i \,\HG_{\mu i}^{~~k}+(\d_i^k \,\partial_\mu -\HG_{\mu i}^{~~k}) s_k &=
0,\cr}
\eqn\ediagonal$$
where $i,k$ are not summed.  (The generalization of these conditions
for massive renormalizable field theories have been derived
in ref.~[\rsonoda].)
The first equation determines a great part of the connection;
in particular, for $\gamma_i \ne \gamma_k$ (which means $i\not= k$)
we obtain
$$\HG_{\mu i}^{~~k} = {H_{\mu i}^{~~k} \d_{s_k,s_i} \over \gamma_i -
\gamma_k}~, \quad {\rm for}~\gamma_i \ne \gamma_k .\eqn\mofconn$$
The second of \ediagonal\ implies
that the connection coefficients relating
two operators of different spins must vanish
$$
\HG_{\mu i}^{~~k} = 0, \quad\hbox{for}\, s_i \not= s_j .\eqn\scons
$$
The equations \ediagonal\ also imply consistency conditions. For example
the first equation shows that for $\gamma_i = \gamma_j$ but $i\not= j$ we
must have $H_{\mu i}^{~~j}=0$. This equation is a necessary condition
generalizing the familiar conditions requiring that the dimension of the
marginal operators should not change under their own flow [\rdijk ]. We
leave a fuller investigation of the consistency conditions for a future
work.

Having obtained $\HG_{\mu i}^{~~k}$ for $\gamma_i \ne \gamma_k$ let us now show
that for $\gamma_i = \gamma_k$ the Virasoro conditions
\freeal\ determine all connection
coefficients in terms of the connection coefficients connecting
primary fields of equal dimensions (assuming we have a unitary theory,
and there are no linear relations in the Verma modules).
Consider the first equation
in \freeal\ for the case that $n$ is a negative integer. Here the choice
of basis we have made will have a significant effect. The first term of
the equation reads $\left( L_{-n} \right)_i^{~k}\, \HG_{\mu k}^{~~j}$.
The sum over states $k$ here only gets a contribution from one state, the
state $k=L_{-n}\, i$, and the matrix element is one. Therefore
$\left( L_{-n} \right)_i^{~k}\, \HG_{\mu k}^{~~j}\equiv\HG_{\mu
L_{-n}i}^{~~j}$,
and we find
$$\HG_{\mu L_{-n}i}^{~~j}
= {1\over 2}\, H_{\mu i}^{~~j}\delta_{s_i,s_j+n}
- \partial_\mu (L_{-n})_i^j
+\sum_k \HG_{\mu i}^{~~k} \, (L_{-n})_k^{~j}.\eqn\arecrel$$
Here in the left hand side we have a descendant field, and
we want to relate this term to connection coefficients where
the lower index represents a primary field. The only case of
interest here is when
$(\Delta_j,\overline\Delta_j) =(\Delta_i+n,\overline\Delta_i)$,
since otherwise the connection coefficient is off diagonal and
therefore known. It follows that under this condition, the
sum indicated in the last term of the above equation can only
run over states $k$ with
$(\Delta_k,\overline\Delta_k) =(\Delta_i,\overline\Delta_i)$ and
we have
$$\HG_{\mu L_{-n}i}^{~~j}
= {1\over 2} H_{\mu i}^{~j}
- \partial_\mu (L_{-n})_i^j
+\sum\limits_{{\Delta_k=\Delta_i}\atop {\overline\Delta_k=\overline\Delta_i}}
\HG_{\mu i}^{~~k}  (L_{-n})_k^{~j}.\eqn\axecrel$$
Note that in the right hand side the connection coefficient involves
the state $i$ and not $L_{-n}i$. Thus this equation can be used
recursively to relate the connection coefficients with an arbitrary
descendant in the lower index, to connection coefficients with a
primary in the lower index. In particular, if $j$ is primary, the sum
vanishes identically.
Note also that in general the states $k$ to
be summed over can be either primary or descendant. This relation therefore
expresses $\HG_{\mu~ desc}^{~~~*}$, with the asterisk denoting
an arbitrary state, in terms of $\HG_{\mu~prim}^{~~prim}$
and $\HG_{\mu~prim}^{~~desc}$.
Thus our problem is now to show that a connection coefficient of the type
$\HG_{\mu~prim}^{~~desc}$ can be found.

To this end consider again Eqn.\freeal\
this time taking $n>0$ and $i$ to be a primary state. We then have
$$\sum_k\HG_{\mu i}^{~~k}(L_n)_k^{~j}=-{1\over 2} H_{\mu i}^{~j}
\delta_{s_i,s_j+n} - \partial_\mu (L_n)_i^j .\eqn\axxrel$$
Since we are interested in diagonal terms we want to implement
the restriction $(\Delta_k,\overline\Delta_k) =(\Delta_i,\overline\Delta_i)$.
In order to attain this we fix $n$ and consider states $j$
such that $(\Delta_j,\overline\Delta_j) =(\Delta_i-n,\overline\Delta_i)$.
Under such circumstances we then find
$$\sum\limits_{{\Delta_k=\Delta_i}\atop {\overline\Delta_k=\overline\Delta_i}}
\HG_{\mu i}^{~~k}(L_n)_k^{~j}=-{1\over 2} H_{\mu i}^{~j}
- \partial_\mu (L_n)_i^j .\eqn\axxrr$$
Note that the sum extends only over $k$'s that must be descendants.
The most general situation one must consider
corresponds to the case when
the primary field $i$ is degenerate with the set of descendants
at level $N$ arising from another primary field $i'$. For the case when
we have a unitary theory, and there are no linear relations in the Verma
module corresponding to the primary field $i'$, it is not hard to see that
the above relations determines all the connection coefficients.
This is done by solving successively for the connection coefficients
$\HG_{\mu i}^{~~k}$, where $k$ are the descendant fields, broken
into groups according to the number of Virasoro operators needed to
obtain them from the primary state:
$L_{-n_1}\ket{i'}$, $L_{-n_1}L_{-n_2}\ket{i'}$ with $n_1\geq n_2$,
$L_{-n_1}L_{-n_2}L_{-n_3}\ket{i'}$ with $n_1\geq n_2 \geq n_3$,
and so on. For each element $L_{-n_1}L_{-n_2}\cdots L_{-n_j}\ket{i'}$,
we take $n=n_1$ in the above formula, and pick
$\ket{j} = L_{-n_2}\cdots L_{-n_j}\ket{i'}$. The interested reader
may verify that this procedure works.

\section{Using Kac-Moody currents}

We have so far discussed one method of setting up a basis and choosing
a set of equations - a method that uses the Virasoro operators.
We will now make a cursory discussion of another possibility which
yields a greater number of coefficients explicitly. Recall that
with the Virasoro method only the coefficients with $\gamma_i \neq \gamma_j$
could be determined explicitly. The full investgation of this possibility
is perhaps best done in the context of an example - a task we leave for the
future.

To motivate the discussion we notice that the equations we solved in the
earlier case arose as the covariant derivative of the Virasoro
operators - or in other words of the conformal field $T(z)$. One might
consider the equations corresponding to the derivatives of other conformal
fields and try to solve the equations in a basis appropriate to those
conformal fields. As discussed earlier the coefficients of a
connection in a particular basis can be determined explicitly only
if that coefficient is zero in any symmetry one form $s_\mu$. We must
choose the conformal fields with care if the number
of explicitly determinable coefficients is to be enlarged.

We now consider the possibility that throughout theory space
we have a set of holomorphic currents $J_a (z)$ (and/or a
set of antiholomorphic ones) generating a Kac-Moody symmetry.
Since a primary operator has dimension $(1,0)$, we have
the usual expansion
$$J_a(z) \equiv \sum_n {J_{a,n}\over z^{n+1}}, \eqn\currexp$$
and the operator product expansion of two currents is given by
$$J_a(z) J_b(w) = {k \delta_{ab} \over (z-w)^2} + {1\over (z-w)}
\, i f_{ab}^c\, J_c (w) + \cdots \eqn\opecurr$$
We now consider evaluating the covariant derivative of the current
operators
$$
D_\mu (\Gamma) J_{a}(z) \equiv D_\mu (\Gamma ) \bigl( \bra{0,z, \infty^*}
J_a\rangle \bigr) =\left( D_\mu (\Gamma) \bra{0,z, \infty^*}
\right) \ket{J_{a}} + \bra{0,z, \infty^*}\left( D_\mu (\Gamma)
\ket{J_{a}} \right), \eqn\ecovarja
$$
where the states $\ket{J_{a}}= J_a (z=0)\ket{0}$ are defined all over theory
space. We must evaluate the right hand side of this equation.
The second term in this right hand side, however, is
ambiguous because $ D_\mu (\Gamma) \ket{J_{a}}$ is ambiguous. We need
to make a choice.  Our choice is not completely arbitrary
since only the connection coefficients relating $\ket{J_a}$ to
the states generated by symmetry transformations on $\ket{J_a}$ are ambiguous.
Then the right hand side is known (at least in principle), and
one can proceed in analogy with the argument that was used to solve for the
connection coefficients in the Virasoro method.
While we do not analyze these equations to solve for the
coefficients we have evaluated the first term of the
right hand side of Eqn.\ecovarja\ for future reference.
We leave this instructive calculation for Appendix A. Using the
connection $\bar c$, we have obtained
$$\bra{\Phi^j}\, \Bigl( D_\mu (\bar c) \bra{0,z,\infty^*} \Bigr)
\hskip-3pt\ket{\Phi_i}\hskip-3pt\ket{J_a}\hskip-3pt
= - {1\over z}\cdot {1\over 2} \, H_{(J_{a,1}\O_\mu ) i}^{~~~~~~~~~j}\,\cdot
\delta_{\Delta_i,\Delta_j}\cdot\delta_{\overline\Delta_i,\overline\Delta_j}.
\eqn\bthatsit$$

\chapter{The Curvature of the Connections}

In this section we will compute the curvature of the connections
we have been studying. The curvature of the connection $c$ has been
already computed in ref. [\rsonoda ]. Here we will streamline the
derivation considerably, and this will allow us to give a simple proof that the
connections $\HG_D$ are flat.

The connection $\HG$ ($\HG_D$ for $D$ equal to the unit disk)
was expected to be flat.
Notice that any
surface state $\bra{\Sigma}$ corresponding to a surface with local
coordinates that cover it precisely must have zero covariant derivative
with respect to the connection $\HG$. Thus we have an infinite
number of linearly independent nonvanishing sections that are
covariantly constant. We would expect zero curvature in analogy to
the situation for
dimension $n$ vector bundles, where the existence of $n$ covariantly
constant basis sections guarantees zero curvature.

\section{Preliminary Discussion}

As we have seen, the general expression for the covariant derivative of
a surface state with an arbitrary connection is of the form
$$D_\mu (\Gamma) \bra{\Sigma} = - \hskip-6pt \int_{\Sigma -
\cup_i D_i}\hskip-6pt d^2 z~ \bra{\Sigma;z} \O_\mu \rangle
-  \sum_{i=1}^n \bra{\Sigma}\omega_\mu^{(i)}.\eqn\exgen$$
We wish to compute the curvature of $\Gamma$ using the above formula.
As we know, the commutator of two covariant derivatives acting on
a tensor must give curvature, that is
$$[ D_\mu (\Gamma ) , D_\nu (\Gamma ) ]\, \bra{\Sigma} =
-\bra{\Sigma} \sum_i \Omega_{\mu\nu}^{~(i)} (\Gamma ) .\eqn\xxcurvaris$$
It is certainly {\it not} manifest that using the right hand side of
Eqn.\exgen\ (the variational formula) to compute the commutator will give
a result compatible with
\xxcurvaris . Our strategy will be to compute the commutator for a particular
connection where it will be relatively straightforward to understand the
conditions under which Eqn.\xxcurvaris , as evaluated with the
variational formula,  holds (and in doing so obtain the
value of the curvature). This will imply that \xxcurvaris\ holds in general,
for we will prove next that upon a change in the connection,
the commutator transforms as expected.

It follows from the formula giving the curvature in terms of the connection
that
$$\Delta_{\mu\nu} \equiv
\Omega_{\mu \nu} (\Gamma + \theta ) - \Omega_{\mu \nu} (\Gamma) =
\Omega_{\mu \nu} (\theta) + \Gamma_\nu \theta_\mu  -\Gamma_\mu \theta_\nu
+\theta_\nu \Gamma_\mu - \theta_\mu \Gamma_\nu.\eqn\echancon$$
We must check if this property follows from our expression for covariant
derivatives of surface states (the variational formula). To do so let us
compute explicitly the additional terms $\Delta_{\mu\nu}$ that arise when
we change the connection from $\Gamma$ to $\Gamma +\theta$. We must verify
that we obtain the expression in the right hand side of \echancon . We find
$$\eqalign{
\Delta_{\mu\nu} &= \left(\, D_\mu (\Gamma + \theta) D_\nu (\Gamma + \theta)
\bra{\Sigma}- D_\mu (\Gamma) D_\nu (\Gamma) \bra{\Sigma} \,\right)
 - \left( \mu\leftrightarrow \nu \right) \cr
&= D_\mu (\Gamma + \theta) \biggl( - \hskip-2pt \int_{\Sigma'}\hskip-2pt
d^2 z~ \bra{\Sigma;z} \O_\nu \rangle -  \sum_{i=1}^n
\bra{\Sigma}(\omega_\nu^{(i)} -\theta_\nu^{(i)} ) \biggr) \cr
& - D_\mu (\Gamma) \biggl( - \hskip-2pt \int_{\Sigma '}\hskip-2pt
d^2 z~ \bra{\Sigma;z} \O_\nu \rangle -  \sum_{i=1}^n
\bra{\Sigma}\omega_\nu^{(i)}\biggr)-(\mu\leftrightarrow\nu ),\cr}\eqn\ssys$$
where $\Sigma' = \Sigma - \cup_iD_i$. Applying the variational formula
again we find
$$\eqalign{
\Delta_{\mu\nu}& = \hskip-2pt \int_{\Sigma '}\hskip-2pt d^2 z~
\sum_{i=1}^{n}
\bra{\Sigma;z} \bigl( \ket{\O_\nu } \,\, \theta_\mu^{(i)}   +
\ket{\O_\mu } \,\, \theta_\nu^{(i)} \bigr) +
\sum_{i,j=1}^n \bra{\Sigma} \bigl( \theta_\mu^{(j)} \omega_\nu^{(i)}+
\omega_\mu^{(j)} \theta_\nu^{(i)}+\theta_\mu^{(i)}\theta_\nu^{(j)} \bigr)\cr
&\qquad - \sum_{i=1}^n \bra{\Sigma}\bigl( D_\mu (\Gamma + \theta)\,
\omega_\nu^{(i)}+ D_\mu (\Gamma + \theta )
\theta_\nu^{(i)}\,  \bigr) - ( \mu \leftrightarrow \nu ) ,\cr }\eqn\noseat$$
and the antisymmetrization leaves us with
$$
\Delta_{\mu\nu} = - \sum_{i=1}^n \bra{\Sigma} \left[ \left( D_\mu (\Gamma)
\theta_\nu^{(i)}  - D_\nu (\Gamma) \theta_\mu^{(i)}\right) + \left(
\theta_\mu^{(i)} \theta_\nu^{(i)}  - \theta_\nu^{(i)}
\theta_\mu^{(i)}\right) \right]. \eqn\erelterm$$
A simple computation expanding the term in brackets in the above equation
shows that, as desired, it is equal to the right hand side of \echancon .
This result establishes that if for some covariant derivative, its commutator,
as computed using the variational formula,
gives just curvature, this will be the case for an
arbitrary covariant derivative.

We would like now to use Eqn.\erelterm\ to investigate what happens to
the curvature if the connection $\Gamma$ is changed by a symmetry, i.e., if
we let  $\theta_\mu = S_\mu$.
We then have$$
\Omega_{\mu \nu} (\Gamma + S) - \Omega_{\mu \nu} (\Gamma) =
 \left( D_\mu (\Gamma) S_\nu  - D_\nu (\Gamma) S_\mu \right) - \left(
S_\mu S_\nu - S_\nu S_\mu \right) .\eqn\esym$$
The second term in Eqn.\esym\ is a commutator of symmetries and therefore
it is itself a symmetry. Let us investigate under what conditions the
first term is also a symmetry. Using $ \sum_{i=1}^{n}\bra{\Sigma}S_{\mu}^{i}=0$
we find
$$
\sum_{i=1}^{n} \bra{\Sigma} \left( D_\mu (\Gamma) S_\nu^{(i)}-D_\nu
(\Gamma) S_\mu^{(i)}  \right)  = -\sum_{i=1}^n \left[ (D_\mu (\Gamma)
\bra{\Sigma})\,\, S_{\nu}^{(i)} - (D_\nu (\Gamma) \bra{\Sigma})\,\,
S_{\mu}^{(i)}
\right] \eqn\hoer$$
We now use the variational formula with some radius $\epsilon$, to obtain
$$\eqalign{
\sum_{i=1}^{n} \bra{\Sigma}  \left( D_\mu (\Gamma) S_{\nu}^{(i)}- D_\nu
(\Gamma) S_{\mu}^{(i)} \right) & = -\sum_{i=1}^{n}
\,\int_{\Sigma '} d^2 z~ \bra{\Sigma;z} \O_{[\mu} \rangle
\,S_{\nu]}^{(i)}-\sum_{i,j=1}^n \bra{\Sigma} \omega_{[\mu}^{(j)}\,
S_{\nu ]}^{(i)}\cr
&=\sum_{i=1}^n \int_{\Sigma '} d^2 z~ \bra{\Sigma;z}S_{[\nu}\ket{\O_{\mu ]}}
+\sum_{i=1}^n \bra{\Sigma}[ S_{[\nu}^{(i)}\, ,\,\omega_{\mu ]}^{(i)}\,].
\cr}\eqn\tder$$
Since $S_{\nu}\ket{\O_\mu} = 0$ (as argued below Eqn.(A.3)), the
vanishing of the right hand side of \tder , for all $\bra{\Sigma}$, requires
that
$$[S_{\nu} \, , \omega_\mu\, ]- [\, S_{\mu}\, ,\, \omega_\nu \,] \approx 0,
\eqn\ecommutator$$
where the equivalence relation $\approx$ means that the objects to the left and
to the right of the symbol are equal up to a symmetry operator.
Eqn.\ecommutator\ is then the condition that ensures that adding a
symmetry to the connection changes the curvature by a symmetry
$$\Omega_{\mu \nu} (\Gamma + S) \approx \Omega_{\mu \nu} (\Gamma) .\eqn\rxsd$$
We see that for certain natural connections the class of
connections related to it by a symmetry all yield identical curvatures
up to symmetries. Eqn.\rxsd\ is a necessary condition on a connection
$\Gamma$ to be able to write its
curvature (up to symmetries) in terms of the pair $(D,\omega_\mu)$. Indeed,
this pair does not change as we vary the connection by a symmetry, implying
thereby that the curvature must not change by more than a symmetry.

\subsection{Verifying Eqn.\rxsd\ for $\HG_D$, $c$ and $\bar c$.}
 The connection $\HG$ satisfies this
naturality condition and so does the connection $c$ and the connection $\bar
c$.
This is readily verified by showing that \ecommutator\ holds.
For $\HG_D$, we can choose a domain $D$ such
that $\omega =0$. For $c$ we choose a radius $\eta$ (which eventually is taken
to zero) and recall that
$$\omega_\mu^c (\eta) = \int_{\eta< |w|<1} \hskip-10pt
d^2w\, \hbox{Ad} \,\bra{0,w,\infty^* }
\O_\mu (w)\rangle ,\eqn\adiagsubt$$
where Ad acting on a matrix operator yields the part of the operator above the
diagonal. To evaluate the left hand side of Eqn.\ecommutator\  we note that
in general for any operator $M = M_{Ad} + M_D +M_{Bd}$, where $Bd$ denotes
the part below the diagonal, and a diagonal operator $S$, we have that
$$
[M_{Ad}, S] = [M,S]_{Ad}, \quad [M_{Bd}, S] = [M,S]_{Bd} .\eqn\splimat$$
For our particular  case let
$$M = \int_{\eta< |w|<1} \hskip-10pt
d^2w\,  \bra{0,w,\infty^* }\O_\mu (w)\rangle ,\eqn\partcase$$
so that $M_{Ad} = \omega_\mu^c (\eta)$. We then find
$$[M,S] =  \int_{\eta< |w|<1} \hskip-10pt d^2w\,\, [\,\bra{0,w,\infty^* }\O_\mu
(w)\rangle \, , S\, ] = - \int_{\eta< |w|<1} \hskip-10pt d^2w\,
\bra{0,w,\infty^* }S \ket{\O_\mu (w)} =0$$
since $S\ket{\O_\mu} = 0$ as discussed above. Back in Eqn.\splimat\ we
find that $[\omega_\mu^c (\eta) , S ] = 0$, which is the desired result.
For the connection $\bar c$
$$\omega_\mu^{\bar c} (\eta) = \int_{\eta< |w|<1} \hskip-10pt
d^2w\, (\hbox{Ad}+\hbox{Bd}) \,\bra{0,w,\infty^* }
\O_\mu (w)\rangle ,\eqn\xadiagsubt$$
and a similar argument using  \splimat\ shows that
$[\omega_\mu^{\bar c} (\eta) , S ] = 0$. This concludes the proof
that the connections $\HG , c$, and $\bar c$ are natural ones.

\section{Computation of $\Omega_{\mu\nu}(c)$ and $\Omega_{\mu\nu}(\bar{c} )$}

Let us now introduce the linear operators $\hbox{Sg}^z$ and  $\hbox{Fp}^z$,
which will be defined acting on functions of $r(=|z|)$ of the following type
$$f_{a,n}(r) ={1\over r^a} \cdot (\ln r)^n \, , \quad n\geq 0 , \eqn\termsdef$$
where $n$ is a constant integer, and where $a=a(x)$ could be a function in
theory space. A simple example of such functions has already appeared
in operator product expansions (see Eqn.\opeex ) where $n=0$, and
$a=2+\gamma_i-\gamma_k$.
If the functions $f_{a,n}(r)$ are integrated around $r=0$ with the usual
measure $rdrd\theta$, the integrals are finite whenever $a<2$, and divergent
whenever $a \geq 2$ (for all values of $n\geq 0$). The operator
$\hbox{Sg}^z$, with Sg for singular, picks the unintegrable functions,
namely, those with $a\geq 2$, and kills the others. The operator
$\hbox{Fp}^z$, with Fp for finite part, picks the integrable functions,
namely, those with $a <2$, and kills
the others. It follows that $\hbox{Sg}^z + \hbox{Fp}^z = 1$, when
acting on sums of functions of the type indicated in \termsdef .
It is an important property that the action of these operators
(on sums of $f$'s) {\it commutes} with the operation of covariant
differentiation we have been studying. This follows because
differentiation does not affect the $r$-dependence
of the functions except in the case when the function $a(x)$ is
differentiated. In this case we pick an extra factor of $\ln r$
and therefore the integrability property of the function is unchanged.

The operator Sg can be used to express the subtraction operator
$\omega_\mu (\eta )$ necessary when taking covariant derivatives
with the connection $c$. We have that
$$\omega_\mu^c (\eta ) = \int_{\eta< |w|<1} \hskip-10pt
d^2w\, \hbox{Sg}^{w} \bra{0,w,\infty^* }
\O_\mu (w)\rangle ,\eqn\recall$$
where, with a slight abuse of notation, we indicate the point where
$\O_\mu$ is inserted in the ket itself.

We consider now a surface state $\bra{\Sigma}$,
which we assume to be equipped with local coordinates at the punctures
such that the associated unit disks are well defined and disjoint.
The covariant derivative $D_\mu (c)$ of such surface states
can be written as
$$D_\mu (c) \bra{\Sigma}= \lim_{\epsilon\to 0}\Bigl[- \hskip-6pt
\int_{|z-y_i|>\epsilon}\hskip-8pt d^2z \bra{\Sigma;z}\O_\mu\rangle
+\sum_i\hskip-3pt
\int_{\epsilon <|z-y_i|< 1}\hskip-10pt d^2z \,\hbox{Sg}^{z-y_i}\bigl(
\bra{\Sigma;z}\O_\mu\rangle \bigr) \Bigr] .\eqn\newexpr$$
We break the first integral into a piece outside the unit disks,
and additional pieces that combine naturally with the second term:
$$D_\mu (c) \bra{\Sigma}= - \hskip-6pt
\int_{|z-y_i|>1}\hskip-8pt d^2z \bra{\Sigma;z}\O_\mu\rangle
-\sum_i\hskip-3pt
\int_{|z-y_i|< 1}\hskip-10pt d^2z \,\hbox{Fp}^{z-y_i}\bigl(
\bra{\Sigma;z}\O_\mu\rangle \bigr) .\eqn\newexpr$$
This is a useful way of writing the covariant derivative $D_\mu (c)$
of a surface section. It should be emphasized, however, that when
we take a second covariant derivative we cannot use this result,
since the surface states obtained after the first derivative do
not satisfy the condition of having disjoint unit disks for {\it all}
punctures. The puncture associated to the insertion of $\O$ ($z$ in
$\bra{\Sigma ;z}$) can be arbitrarily close to the other punctures.

We can now begin our computation.
Making use of \newexpr\ we immediately obtain the following expression for
the commutator of two covariant derivatives
$$ [\, D_\mu (c) , \, D_\nu (c)\, ]\, \bra{\Sigma} =
-\hskip-6pt
\int_{|z-y_i|>1}\hskip-8pt d^2z F_{\mu\nu}(z)
-\sum_i\hskip-3pt
\int_{|z-y_i|< 1}\hskip-10pt d^2z \,\hbox{Fp}^{z-y_i}\bigl(
F_{\mu\nu}(z)\bigr) ,\eqn\curvpr$$
where $F_{\mu\nu}(z)$ is given by
$$\eqalign{
F_{\mu\nu}(z) &\equiv \, D_\mu (c) \bigl( \bra{\Sigma ;z}\O_\nu\rangle \bigr)
- D_\nu (c) \bigl( \bra{\Sigma ;z}\O_\mu\rangle \bigr) \cr
&= \lim_{\eta\to 0}\Bigl[ -\hskip-6pt\int_{|w-y_i|>\eta \atop |w-z|>\eta}
\hskip-10pt d^2w \bra{\Sigma ;z,w}\bigl(\ket{\O_\mu (w)}\ket{\O_\nu (z)}-
\ket{\O_\nu (w)}\ket{\O_\mu (z)}\bigr)\cr
&\quad\quad +\bra{\Sigma ;z}
\bigl( \omega_\mu^c(\eta ) + \sum_i \omega_\mu^{c(i)}(\eta )
\bigr) \ket{\O_\nu (z)} \cr
&\quad\quad -\bra{\Sigma ;z}\bigl( \omega_\nu^c(\eta )
+ \sum_i \omega_\nu^{c(i)}(\eta )\bigr) \ket{\O_\mu (z)}\Bigr] \cr
&\quad\quad + \bra{\Sigma ;z}
\bigl( D_\mu \ket{\O_\nu}-D_\nu \ket{\O_\mu}\bigr) . \cr}\eqn\hummm$$
Here $\omega_\mu^c$ is defined in \adiagsubt , and, following our comments
above, the derivative of the
surface states $\bra{\Sigma ;z}$ has been computed using the original
expression and not the simplified form in \newexpr .

\subsection{Analysis of Singularities}
In order to simplify further our analysis we must understand the nature
of the singularities as the operators $\O_\mu$ and $\O_\nu$ get close.
For simplicity we choose our basis of states in a way that the marginal
operators are basis states throughout theory space. In a unitary theory
we must have the following type of OPE
$$\O_\mu (z) \O_\nu (0) = {G_{\mu\nu}\over r^4}\, {\bf 1} + \sum_{\gamma_k>0}
{H_{\mu\nu}^{~~k} \over z^{2-\Delta_k} {\bar z}^{2-\bar \Delta_k}} \,\Phi_k(0).
\eqn\tmarcl$$
Making use of this equation and translational invariance we can derive the
operator expansion of $\O_\nu (z) \O_\mu (0)$ and obtain for the
antisymmetrized
combination the following result
$$\eqalign{\O_{[\mu} (z) \O_{\nu ]}(0) &= \sum_{\gamma_p>0}
{H_{\mu\nu}^{~~p}- H_{\nu\mu}^{~~p}\over z^{2-\Delta_p}
{\bar z}^{2-\bar\Delta_p}} \,\Phi_p(0)\cr
&= -\sum_{\gamma_k>0}
H_{\mu\nu}^{~~k}(-1)^{\gamma_k}\biggl(
{\partial \Phi_k(0) \over z^{1-\Delta_k} {\bar z}^{2-\bar \Delta_k} }
+{\bar\partial \Phi_k(0) \over z^{2-\Delta_k} {\bar z}^{1-\bar \Delta_k}}
+\cdots \biggr) \cr}\eqn\zsaw$$
where the dots indicate terms with two or more derivatives. Now consider
integrating over $z$ in a disk surrounding $z=0$. All terms with two
or more derivatives give finite
contributions. The first term can only give a divergence if $\gamma_k\leq 1$.
Since the angular integration forces $s_k = -1$ to get a nonvanishing answer,
the field $\Phi_k$ must be a $(0,1)$ field, and therefore purely
antiholomorphic. Therefore $\partial\Phi_k$ vanishes and we cannot get any
divergence from the first term.
Exactly the same argument applies for the second term. Thus, we conclude that
the antisymmetric combination in the left hand side of Eqn.\zsaw\ can be
integrated over $z$ without the need for subtractions. This implies that
the integral over $w$ in \hummm\ does not require the condition
$|w-z|>\eta$. Therefore the explicit
infinite subtraction terms must also vanish
$$\bra{\Sigma ;z} \bigl( \omega_\mu^c (\eta ) \ket{\O_\nu (z)} -
\omega_\nu^c (\eta ) \ket{\O_\mu (z)}\,\bigr) = 0,\eqn\vansubs$$
as can be verified by an argument completely analogous to the one given
before.

\subsection{Analysis of Torsion}  We must now consider the terms
$$ D_\mu \ket{\O_\nu}-D_\nu \ket{\O_\mu} = (c_{\mu\nu}^{~~k} -
c_{\nu\mu}^{~~k})
\ket{\Phi_k}, \eqn\iftor$$
where, in writing the right hand side, we used a basis where the marginal
operators are some of the basis vectors.
The sum over $k$ can only extend over operators of spin zero and of total
dimension less than or equal to two (recall $c$ is upper triangular). Thus
the above breaks into two type of terms
$$ D_\mu \ket{\O_\nu}-D_\nu \ket{\O_\mu} =
\sum_\rho (c_{\mu\nu}^{~~\rho} - c_{\nu\mu}^{~~\rho})
\ket{\O_\rho}+
\sum_{\gamma_p <2} (c_{\mu\nu}^{~~p} - c_{\nu\mu}^{~~p})
\ket{\Phi_p},  \eqn\ifxtor$$
where we recall that the dimension $(1,1)$ fields must be primary. Making
use of \relconni\ and \trmofconn , we find
$$c_{\mu \nu}^{~~p}-c_{\nu\mu}^{~~p} = {(H_{\mu \nu}^{~~p}-H_{\nu\mu}^{~~p})
\over 2-\gamma_p} \,\delta_{s_p,0}
, \quad\hbox{for} \, \gamma_p < 2 .\eqn\soxffar$$
We can use Eqn.~\zsaw\ to show that this vanishes. Using the first term
in the expansion of that equation, we see that $\Delta_p = \Delta_k+1>1$.
Since any field $\Phi_p$ entering in \soxffar\ must be of spin zero this
implies that $\gamma_p >2$, and therefore no such field contributes. An
exactly analogous argument holds for all other terms in \zsaw .
Thus indeed $c_{\mu \nu}^{~~p}-c_{\nu\mu}^{~~p}$ vanishes for $\gamma_p<2$
and $s_p = 0$. Back in \ifxtor\ we then have
$$ D_\mu \ket{\O_\nu}-D_\nu \ket{\O_\mu} =
\sum_\rho (c_{\mu\nu}^{~~\rho} - c_{\nu\mu}^{~~\rho})
\ket{\O_\rho}=\sum_\rho T_{\mu\nu}^{~~\rho}\ket{\O_\rho},\eqn\torifs$$
where we introduced the torsion-like object $T_{\mu\nu}^{~~\rho}$.
Unless $T_{\mu\nu}^{~~\rho}$ vanishes, the term
$D_\mu \ket{\O_\nu}-D_\nu \ket{\O_\mu}$ present
in $F_{\mu\nu}$ would end up, back in \curvpr , giving us a result of the
form $[D_\mu , D_\nu ] \bra{\Sigma} = -\bra{\Sigma}\Omega_{\mu\nu} +
T_{\mu\nu}^{~~\rho}D_\rho \bra{\Sigma}$. This would be inconsistent since
on a vector bundle the commutator of two covariant derivatives must only give
curvature and cannot include true torsion terms. We must therefore have that
$$T_{\mu\nu}^{~~\rho}=c_{\mu\nu}^{~~\rho}-c_{\nu\mu}^{~~\rho}=0.\eqn\tconstr$$
This condition is indeed satisfied for the connection $\HG$ in the particular
case of toroidal compactification. We can give an argument that suggests
that indeed the condition is satisfied on general grounds. Consider the
partition function $Z_\Sigma$ on the surface $\Sigma$ (which must have
no punctures) as a function in theory
space. If we consider the commutator $[D_\mu , D_\nu ]$ acting
on $Z_\Sigma$ we must get zero because $D_\mu D_\nu Z_\Sigma\equiv \partial_\mu
\partial_\nu Z_\Sigma$, by definition. Since the curvature term cannot act on
a function we obtain that $T_{\mu\nu}^{~~\rho} D_\rho Z_\Sigma = 0$. But
$D_\rho Z_\Sigma = \langle\langle \O_\rho \rangle\rangle_\Sigma$, where the
right hand side denotes expectation value of the operator as integrated
all over the surface $\Sigma$.\foot{This can be shown by first
writing $Z_\Sigma = \bra{\Sigma '}0\rangle$, where $\Sigma '$ is a one
punctured surface with the same conformal structure as $\Sigma$ if we
forget about the extra puncture. Taking a covariant derivative with
the connection $\HG$ one can see that the result amounts to integrating
the insertion over the whole of the unpunctured surface.} Therefore
we know that
$$T_{\mu\nu}^{~~\rho} \,\, \langle\langle \O_\rho \rangle\rangle_\Sigma = 0.
\eqn\cansay$$
This is an equation for every unpunctured surface $\Sigma$, and it suggests
strongly that $T_{\mu\nu}^{~~\rho} = 0$. The expectation value of the
marginal fields cannot vanish in general for genus $g\geq 1$ since otherwise
the partition function would be a constant in theory space. We will not
attempt to prove on general grounds the vanishing of the torsion $T$, this
would
involve showing that the expectation values of the marginals, thought as
defining a vector in the tangent space to theory space, span this tangent
space as we vary the surfaces. We expect this must be the case. A similar
argument was made in ref. [\rsonoda ] to prove that $T_{\mu\nu}^{~~\rho}=0$
for the case of massive renormalizable theories .

\subsection{Completing the Computation}
All in all, our discussion above implies that $F_{\mu\nu}$ in \hummm\
reduces to the following expression
$$F_{\mu\nu}(z)
= \lim_{\eta\to 0}\Bigl[\, -\hskip-8pt\int_{|w-y_i|>\eta}
\hskip-10pt d^2w \bra{\Sigma ;z,w}\bigl(\ket{\O_{[\mu} (w)}
\ket{\O_{\nu ]}(z)}
\bigr) +\sum_i \bra{\Sigma ;z}
\omega_{[\mu}^{(i)}(\eta ) \ket{\O_{\nu ]} (z)} \Bigr] .\eqn\nhummm$$
We now split the integral in Eqn.\nhummm\ into a piece where $w$
is outside the unit disks, and pieces where $w$ is inside the various
$y_i$ disks. For the latter pieces we define $w_i = w-y_i$. We then obtain
$$\eqalign{F_{\mu\nu}(z)
&=\, -\hskip-6pt\int_{|w-y_i|>1}
\hskip-10pt d^2w \bra{\Sigma ;z,w}
\bigl(\ket{\O_{[\mu } (w)}\ket{\O_{\nu ]}(z)}\bigr)\cr
&-\sum_i\hskip-2pt\int_{|w_i| < 1}
\hskip-10pt d^2w_i \biggl( \bra{\Sigma ;z,w_i}- \bra{\Sigma ;z}
\hbox{Sg}^{w_i} \bra{0,w_i, \infty^{*(i)}} \biggr)
\ket{\O_{[\mu } (w_i)}\ket{\O_{\nu ]}(z)} .\cr}\eqn\xhummm$$
In writing this expression we have dropped the $\eta \to 0$ limit
since the integrand, as written, is integrable around $w_i=0$.
We cannot simplify further the expression in parenthesis in the
second line of the above equation because the
location of $z$ is arbitrary. The rest of the computation of curvature
does not involve any conceptual difficulty and we have therefore
relegated the details to Appendix B.1. The result is
$$\Omega_{\mu\nu} = \int_{|z|<1}
\hskip-8pt d^2z \, \hbox{Fp}^z \hskip-8pt\int_{|w| < 1}
\hskip-8pt d^2w \biggl( \bra{0 ,z,w, \infty^*} - \bra{0^k,z,\infty^*}
\hbox{Sg}^w \bra{0,w, \infty^{*k} } \biggr)
\ket{\O_{[\mu }}\ket{\O_{\nu ]}} ,\eqn\result$$
where the $0^k$ and $\infty^k$ state spaces are contracted. Here
the curvature is expressed as a double integral over four point
functions [\rsonoda ].  We can use this formula to obtain an expression
for the curvature  in terms of OPE coefficients. The computation is given
in detail in Appendix B.2, and the result is
$$
\Omega_{\mu\nu i}^{~~~j}  = \delta_{s_i,s_j}
\biggl( \,\sum_{\gamma_k > \gamma_i} + \sum_{\gamma_k < \gamma_j }\,\biggr)
\, {H_{[\mu i}^{~~k} H_{\nu ] k}^{~~j} \,\delta_{s_k,s_i} \over
\gamma_{kj}\, \gamma_{ki}}\, , \;\;\;\;
\hbox{for} \,\,\gamma_i \geq \gamma_j ,\eqn\xcurvcfin$$
and $\Omega_{\mu\nu i}^{~~~j}  = 0$, for $\gamma_i < \gamma_j$.

Once the curvature of $c$ is known we also know the curvature of $\bar c$.
It follows directly from the definition of curvature that the diagonal
part of the curvature of an upper triangular connection equals the
curvature of the diagonal part of the connection. This means that
$$\Omega_{\mu \nu} (\bar{c}) = \hbox{Diag} \,\,(\Omega_{\mu \nu} (c)).
\eqn\curvcb$$

\section{Curvature of the Connection $\HG_D$}

We will now show that the connection $\HG_D$ is flat.
We first relate the
connection $\HG_D$ with domain $D$ to the connection $\HG$.
We have that
$$\HG_D = \HG + \omega^D , \quad \hbox{with}\quad
\omega^D_\mu = \int_{\Delta D}
d^2z \bra{0,z,\infty^*} \O_\mu (z)\rangle,\eqn\relatehg$$
where $\Delta D = D^1 - D$ and $D^1$ is the unit disk.
Moreover $\HG=c-\widehat \omega$, where  $\widehat \omega_\mu$ is
given by
$$ \widehat \omega_\mu = \int_{|w|<1} \hskip-10pt
d^2w\, \hbox{Fp}^{w} \bra{0,w,\infty^* }
\O_\mu (w)\rangle .\eqn\recallp$$
All this implies that
$$D_\mu (\HG_D ) \bra{\Sigma} = D_\mu (c)\bra{\Sigma} + \bra{\Sigma}
\sum_i \bigl( \widehat \omega_\mu^{(i)}-\omega^{D(i)}_\mu \bigr)
\eqn\relder$$
Starting from
$$D_\nu (\HG_D ) \bra{\Sigma}= - \hskip-6pt
\int_{\Sigma-D^i}\hskip-8pt d^2z \bra{\Sigma;z}\O_\nu\rangle ,
\eqn\beghere$$
we find
$$ [\, D_\mu (\HG_D ) , \, D_\nu (\HG_D )\, ]\, \bra{\Sigma} =
-\hskip-6pt
\int_{\Sigma - D^i}\hskip-8pt d^2z F_{\mu\nu}(z)
-\sum_i\hskip-3pt
\int_{\Sigma -D^i}\hskip-10pt d^2z \bra{\Sigma ;z}
\bigl( \widehat \omega_{[\mu}^{(i)} - \omega^{D(i)}_{[\mu } \bigr)
\ket{\O_{\nu ]}(z)}  .\eqn\ghier$$
The second term in the right hand side, with the help of
\recallp\ and \relatehg\ can
be written as
$$\eqalign{
&-\int_{\Sigma - D^i}
\hskip-8pt d^2z \, \bra{\Sigma ; z} \sum_i \hskip-5pt\int_{|w_i| < 1}
\hskip-10pt d^2w_i \, \hbox{Fp}^{w_i} \bigl( \bra{0,w_i, \infty^{*(i)}}\bigr)
\ket{\O_{[\mu } (w_i)}\ket{\O_{\nu ]}(z)}\cr
&+\int_{\Sigma - D^i}
\hskip-8pt d^2z \, \bra{\Sigma ; z} \sum_i \hskip-2pt\int_{\Delta D^i}
\hskip-7pt d^2w_i \,  \bra{0,w_i, \infty^{*(i)}}
\O_{[\mu } (w_i)\rangle\ket{\O_{\nu ]}(z)}\, .\cr}\eqn\xfirstpart$$
The first term in the right hand side of \ghier\ is evaluated
by taking the expression for $F_{\mu\nu}$ in \xhummm\
and splitting its first term into integrals over $\Sigma - D^i$ and
integrals over $\Delta D^i$. Substituting back in \ghier , and
with a little work, we find that the right hand side of \ghier\
reduces to
$$
\int_{\Sigma - D^i}\hskip-8pt d^2z \sum_i
\hskip-2pt\int_{D^i }
\hskip-1pt d^2w_i \biggl( \bra{\Sigma ;z,w_i}- \bra{\Sigma ;z}
\bra{0,w_i, \infty^{*(i)}} \biggr)
\ket{\O_{[\mu } (w_i)}\ket{\O_{\nu ]}(z)} = 0,\eqn\ufff$$
which vanishes because $z$ must be outside the disks $D^i$
and $w$ must be inside the disks.
Under these circumstances the bra in parenthesis vanishes
identically.
This proves the desired result, namely, we obtain that
$\Omega (\HG_D\, )_{\mu\nu} \approx 0$.

\chapter{Building a Theory in the State Space of Another}

As we have discussed earlier, building the theory at $x$ in the state space of
a theory at $x'$ requires that for every surface state $\bra{\Sigma (x)}$
we should find an associated state in $\H_{x'}$ such that the algebra of
sewing corresponding to the theory at $x$ is respected. We have also seen that
this can be done by simply transporting  the states $\bra{\Sigma (x)}$
from the state space $\H_x$ to the space $\H_{x'}$. A priori, this transport
can be done with {\it any} connection. Even a zero connection may be used.
The drawback, however, is that with this choice we cannot write simple or
natural expressions for the transported state.  In general, the result
of the transport will depend on the path chosen due to the curvature of the
connection. When we transport the surface states we must keep the path
fixed. Then it follows that using different paths simply corresponds to two
representations of the theory differing by a similarity transformation.
Thus connections with curvature are perfectly sensible.

As we have shown in the previous section, the connection $\HG_\mu$ is
flat, in addition to metric compatible. Thus it is a particularly natural
candidate for parallel transport.
We will show, however, that one cannot integrate the equations of parallel
transport for this connection. This, of course, happens because we are
working in an infinite dimensional vector bundle. Therefore, if we wished
to use the connection $\HG$ beyond first order, we must regulate the
infinities. Rather than working with infinite quantities, we can use
connections
that do not have this integrability problem. As we will discuss, the connection
$c$ or the connection $\bar c$ are suitable for finite distance parallel
transport. A state will be called finite if all of its components are finite.

\section{Second Order Nonintegrability of $\HG$}

Let us consider the parallel transport of ket-states by a finite amount.
We choose a one-parameter
family of theories $x^\mu (s), s \in [0,\epsilon]$.  Then we can
do parallel transport using the path
ordered integral of the connection $\HG_\mu$ along the
path from $s=0$ to $s=\epsilon$ using the formula given in Eqn.\eintegral .
The second order term, as we will see now, is ill-defined due to the
divergence of the product of two connections.
For $\HG_i = \HG_j$ and $s_i = s_j$, we can split the sum over
intermediate states into two parts:
$$
(\HG_\mu \HG_\nu)_{i,j} =
\sum_k \HG_{\mu i}^{~~k} \HG_{\nu k,j}
= \left(\sum_{\gamma_k > \gamma_i}
+ \sum_{\gamma_k  \le \gamma_i} \right)
\HG_{\mu i}^{~~k} \HG_{\nu k,j} .\eqn\esum
$$
Now, the second sum, involving a finite number of terms, is finite.
The first sum is rewritten as
$$\sum_{\gamma_k > \gamma_i}
\HG_{\mu i}^{~~k} \HG_{\nu k,j}
 = - \int_{|z'|< 1} {d^2 z' \over |z'|^2}~ F(z') ,\eqn\split$$
where $F(z')$ is given by
$$F(z') =
\int_{|z|< |z'|} d^2 z
\quad \vev{\O_\nu (1) \biggl( \O_\mu (z) \Phi_i (0)
- {1 \over 2 \pi} \sum_{\gamma_k \le \gamma_i} {D_{\mu i}^{~~k}\delta_{s_i,s_k}
\over |z|^{2+\gamma_i - \gamma_k}}~ \Phi_k (0)
\biggr) \Phi_j (\infty)} .\eqn\ediv
$$
This formula can be verified  by making use of Eqns.\opeex , \mofconn ,
and the assumption that $\gamma_i=\gamma_j$.
The integral in \split\ turns out to be divergent due to the singularity
$$
\O_\mu (z) \O_\nu (z')\simeq {G_{\mu\nu} \over |z-z'|^4} ~{\bf 1} + \cdots ,
\eqn\esing$$
where $G_{\mu\nu} \equiv \vev{\O_\mu\mid\O_\nu}$.
In fact, using \esing\ one can check that $F(z') \sim (1-|z'|)^{-2}$ as
$z'\to 1$, which leads to a divergence when integrated as in \split .
This confirms that at second order the integration with
the connection $\HG_\mu$ fails.
Since the variation of the correlators must be finite,
the above divergences, if kept track properly, ought to be canceled by
the divergences in the second order parallel transport of the surface
states. We will not try to implement this since our aim is
to give a construction manifestly free of divergences.

\section{Integrability of $c$ and $\bar c$}

We will see now that the difficulties observed in the previous subsection
dissappear for the connections $c$ or $\bar c$. We first argue that the
connection elements themselves are finite (this is also true for $\HG$).
This is clear from the variational formulas applied to surface states; the
right hand side of the variational formula is always finite, and therefore
the connection coefficients,
which are determined up to symmetries, can be chosen to be finite.
Consider again paralell
transport of ket-states. It follows from the lower triangular property of
$c$ that in
$$ D_\mu (c) \ket{\Phi_i} = c_{\mu i}^{~~k}\, \ket{\Phi_k},\eqn\finconn$$
the sum in the right hand side involves a finite number of states, those
whose dimensions are lower than or equal to $\gamma_i$. It is then clear
that multiple
covariant derivatives $D_{\mu_1} \cdots D_{\mu_n} \ket{\Phi_i}$ must
be always a finite state with dimension less than or equal to $\gamma_i$.
This proves that finite distance parallel transport of any ket state of
definite dimension is well defined. Note, however, that for a general
state $s^i\ket{\Phi_i}$ paralell transport with $c$ may not be well defined
if this sum involves an infinite number of states. This happens because in
$D_\mu (s^i\ket{\Phi_i})$ involves the term  $s^i c_{\mu i}^{~~k}\ket{\Phi_k}$,
and for fixed $k$ the sum over $i$ can run over an infinite set of values.

Since our description of correlators has been given using the contraction
of bras representing surface states against kets corresponding to local
operators, and having already checked that the transport of the kets is
not problematic, let us now discuss the transport of the bras. The equation
$$ D_\mu (c) \bra{\Phi^i} = -c_{\mu k}^{~~i}\,\,\, \bra{\Phi^k},\eqn\finbra$$
shows that the parallel transport of a basis bra gives a bra with infinite
number of components all with dimension greater than or equal to $\gamma_i$.
The covariant derivative of an {\it arbitrary} finite bra
section must be finite because the resulting component along
any basis bra $\bra{\Phi^k}$ can only get contributions
from bras of equal or lower
dimensionality, and there are only a finite number of such bras. It follows
that the covariant derivative with $c$ of surface sections are always finite
sections. Therefore multiple covariant derivatives are well defined as well,
and as a consequence we can define finite distance paralell transport of
surface states using the connection $c$. All of our arguments for $c$
apply to $\bar c$ as well. For the connection $\bar c$, however, even the
parallel transport of arbitrary finite kets is well defined.

\chapter{Conclusions}

In this paper we have associated a pair ($D, \omega_\mu$) with
any covariant derivative by examining its action on the surface
sections. We have identified a natural class of connections
$\HG_D$ for which $\omega_\mu$ can be set to zero. The connections
$\HG_D$ have zero curvature and $\HG$ (for $D=D^1$, the unit disc)
is metric compatible. $\HG$ does not lead to perturbatively
finite parallel transport. Another class of
natural connections have $\omega_\mu \not= 0$ for any choice of
excluded domain around the punctures. The $\omega_\mu$'s, however,
are constructed in terms of surface states. Among these connections
we have studied the connections $c$ and $\bar c$, which turn out
to be the upper triangular and diagonal parts of $\HG$. Both have
nonzero curvature, and $\bar c$ is compatible with the metric. Both
lead to perturbatively finite parallel transport. For these
connections we were able to obtain expressions for curvature
in terms of the states of the theory, $D$ and $\omega_\mu$.
On general grounds this is not possible unless $\omega_\mu$ satisfies
Eqn.\ecommutator .

We now indicate some directions of future work that emerge from the present
paper. Though we have identified some interesting connections, the notion
of a natural connection has not been made fully precise.
We have also not explored completely
the consistency conditions that follow from the variational formula
\xegen\ on the correlation functions of the theories.
It may be of interest to elucidate the relation between our work and
the work on deformations of topological field theories [\rcecottivafa ].
Finally, a crucial issue is extending our present work to more general
spaces of two dimensional field theories.

\ack{H. Sonoda acknowledges the hospitality of the Center for Theoretical
Physics at MIT.}

\APPENDIX{A}{A: \ Computation of Kac-Moody currents}

We want to compute the right hand side of Eqn. \ecovarja . This is
$$
\left( D_\mu (\Gamma) \bra{0,z, \infty^*}
\right) \ket{J_{a}} + \bra{0,z, \infty^*}\left( D_\mu (\Gamma)
\ket{J_{a}} \right) \eqn\ecovarja
$$
The second of term involves no computation since it is known once
$\left( D_\mu (\Gamma) \ket{J_{a}} \right)$ is fixed. We need to evaluate
the first term $ \left( D_\mu (\Gamma) \bra{0,z, \infty^*} \right)
\ket{J_{a}}$.

\subsection{Preliminaries}
Since we will assume throughout that we have a unitary conformal
field theory, the operator product expansion of the dimension $(1,0)$ current
with the dimension $(1,1)$ marginal operator $\O_\mu$ must be of
the form
$$ J_a(z) \O_\mu (w,\bar w) = {1\over (z-w)^2} (J_{a,1}\O_\mu )(w,\bar w)
+ {1\over (z-w)} (J_{a,0}\O_\mu ) (w,\bar w ) + \hbox{regular}, \eqn\expope$$
since the lowest dimensional operator must be of dimension zero.
We will assume that
$$ J_{a,0} \O_\mu \equiv 0 , \eqn\musttake$$
since this operator, if not zero, must be a $(1,1)$ primary
(recall that $[L_n , J_{a,0}] = 0$). Then the charges associated
to the Kac-Moody symmetry would rotate the marginal operator $\O_\mu$
into some other marginal operator $\O$. All marginal operators
belonging to the same multiplet would generate equivalent deformations
throughout theory space. We do not wish to consider here such degenerate
situation. The operator $(J_{a,1}\O_\mu )(w,\bar w)$ entering in the above
expansion must be a dimension $(0,1)$ field, therefore it must
be primary and and purely antiholomorphic. All in all our operator
product expansion will read
$$ J_a(z) \O_\mu (w,\bar w) = {1\over (z-w)^2} (J_{a,1}\O_\mu )(\bar w)
+ \hbox{regular}, \eqn\opeoj$$
We also claim that the subtraction matrix $\D_\mu (\e )$ vanishes
for currents, that is
$$ \D_{\mu\, J_a}^{~~~~k} (\e ) \equiv 0 .\eqn\yesvanish$$
This follows from \expsub ; if $\gamma_k < 1$, we get zero
because there is no field of spin one of dimension less than
one; if $\gamma_k = s_k =1$ the field $\Phi_k$ must be another
holomorphic current. Nevertheless such an operator cannot
appear in the operator product expansion of an $\O_\mu$ and
a holomorphic current (see \opeoj ).

\subsection{The computation.}
Let us now begin our computation. It follows from \ppxyy\
(with the limit as $\e \to 0$ implicit)
$$\eqalign{
\bigl( D_\mu (\bar c) \bra{\Sigma (0,z,\infty)} \bigr)
\hskip-3pt\ket{\Phi_i}\hskip-3pt\ket{J_a}\hskip-3pt\ket{\Phi_j}
=& - \hskip-10pt \int\limits_{{\e \leq |w| \leq 1/\e}
\atop {|w-z|\geq \e }}\hskip-10pt d^2w \,
\VEV{\O_\mu (w) \Phi_i (0) J_a (z) \Phi_j (\infty )} \cr
&+ \D_{\mu j}^{~~k} (\e ) \VEV{\Phi_i(0) J_a(z) \Phi_k(\infty ) }\cr
& + \D_{\mu i}^{~~k} (\e ) \VEV{\Phi_k(0) J_a(z) \Phi_j(\infty ) },\cr}
\eqn\bigmess$$
where use was made of Eqns.\ppxyy\ and \yesvanish . The last term
in the above right hand side, with the help of the operator product
expansion Eqn. \opeoj , can be written as
$$
-\sum_n {1\over z^{n+1}} \D_{\mu i , J_{a,-n}j} (\e )
=-\sum_{n\leq -1} {1\over z^{n+1}} \D_{\mu i , J_{a,-n}j} (\e )
+\sum_{n\geq 0} {1\over z^{n+1}} \D_{\mu J_{a,n}i ,j} (\e ),\eqn\reareq$$
where in the last step we broke the sum into two pieces and used
the identity discussed in Calculation 3 at the end of this appendix.
We now assume for convenience that $|z| < 1$. Then the integral
above can be split into the region $|w| \leq 1$ and the region
$|w| \geq 1$. For the first piece we use the Ward identity detailed
in Calculation 1 at the end of this appendix
and the second piece is left as it is. This together with \reareq\
enables us to find that the right hand side of \bigmess , henceforth
called $(I)$, becomes
$$\eqalign{
(I)= &-\int\limits_{{\e \leq |w| \leq 1} \atop {|w-z|\geq \e } }
\hskip-6pt d^2w {\partial \over \partial w} \Bigl\{ {1 \over {z-w}}
\VEV{(J_{a,1}\O_\mu ) (\bar w) \Phi_i(0) \Phi_j(\infty ) } \Bigr\}
\cr
&+\sum_{n\leq -1} {1\over z^{n+1}}
\Bigl(\hskip-6pt \int_{\e \leq |w|<1} \hskip-6pt d^2w
\VEV{\O_\mu (w) \Phi_i(0) (J_{a,-n}\Phi_j)(\infty ) }
-\D_{\mu i, J_{a,-n}j}(\e)\Bigr)
\cr
&-\sum_{n\geq 0} {1\over z^{n+1}}
\Bigl(\hskip-6pt \int_{\e \leq |w|<1} \hskip-6pt d^2w
\VEV{\O_\mu (w) (J_{a,n}\Phi_i)(0)\Phi_j(\infty ) }
-\D_{\mu J_{a,n}i,j}(\e )\Bigr)
\cr
&- \hskip-8pt \int\limits_{1\leq |w| \leq 1/\e}
\hskip-10pt d^2w \, \VEV{\O_\mu (w) \Phi_i (0) J_a (z) \Phi_j (\infty )}
+ \sum_n {1\over z^{n+1}} \D_{\mu j,J_{a,n}i} (\e ) \cr}\eqn\getmess$$
The last line in the above expression is evaluated by taking the
operator product expansion of $J_a$ and $\Phi_i$, the
operator product expansion of $\O_\mu$ with $\Phi_j$ and integrating,
the second term subtracting away divergent pieces. This calculation
is quite similar to that leading to the equation
$$\int_{\e \leq |w|\leq 1} \hskip -6pt
d^2 w \,\VEV{\O_\mu (w) \Phi_i(0) \Phi_j(\infty)}
-\D_{\mu i,j} (\e) =\, - \,{ H_{\mu i,j}\over \gamma_i-\gamma_j}
\,\delta_{s_i,s_j}\,(1-\delta_{\gamma_i,\gamma_j}),\eqn\substract$$
which holds for arbitrary operators $\Phi_i$ and $\Phi_j$ (note that
when $\gamma_i=\gamma_j$ we take the right hand side to be zero).
Eqn.\substract\ is also used
to simplify the second and third lines in \getmess .
Using $H_{\mu i,j} = H_{\mu j,i}$, which holds for arbitrary $i,j$,
one finds
$$\eqalign{
(I)= &-\hskip-4pt\int\limits_{{\e \leq |w| \leq 1} \atop {|w-z|\geq \e } }
\hskip-6pt d^2w {\partial \over \partial w} \Bigl\{ {1 \over {z-w}}
\VEV{(J_{a,1}\O_\mu ) (\bar w) \Phi_i(0) \Phi_j(\infty ) } \Bigr\}
\cr
&-\sum_{n\leq -1} {1\over z^{n+1}}\,
{H_{\mu j,J_{a,n}i} + H_{\mu J_{a,-n}j,i} \over \gamma_i-\gamma_j-n}
\,\delta_{s_i-s_j,n}\, (1-\delta_{\gamma_i-\gamma_j, n}).\cr}\eqn\sofarsogood$$
We can now use the result of Calculation 2 at the end of this Appendix to find
$$\eqalign{
(I)= &-\hskip-4pt\int\limits_{{\e \leq |w| \leq 1} \atop {|w-z|\geq \e } }
\hskip-6pt d^2w {\partial \over \partial w} \Bigl\{ {1 \over {z-w}}
\VEV{(J_{a,1}\O_\mu ) (\bar w) \Phi_i(0) \Phi_j(\infty ) } \Bigr\}
\cr
&-{1\over 2}\sum_{n\leq -1} {1\over z^{n+1}}
H_{(J_{a,1}\O_\mu ) i,j}\, \delta_{\Delta_i,\Delta_j}\,
\delta_{\overline\Delta_i+n,\overline\Delta_j}.\cr}\eqn\lookinggood$$
We finally evaluate the integral above by using Stokes theorem.
We get no contribution from $w\to z$, the contribution from
$|w|=1$ cancels precisely the last line of \lookinggood , and
the contribution from $w\to 0$ gives us the final result
$$\bigl( D_\mu (\bar c) \bra{\Sigma (0,z,\infty)} \bigr)
\hskip-3pt\ket{\Phi_i}\hskip-3pt\ket{J_a}\hskip-3pt\ket{\Phi_j}
= - {1\over z}\cdot {1\over 2} \, H_{(J_{a,1}\O_\mu ) i,j}\,\cdot
\delta_{\Delta_i,\Delta_j}\cdot\delta_{\overline\Delta_i,\overline\Delta_j}.
\eqn\thatsit$$
This is our final result for the computation.

\subsection{Calculation 1.}
Our first objective is to find an expression for the one-form
meromorphic object
$$ \omega = \VEV{ \O_\mu (w) \Phi_i(0) J_a (z) \Phi_j(\infty )} \,dz ,
\eqn\formev$$
where the correlator is on the sphere with uniformizing coordinate
$z$ and local coordinates $z$ around $z=0$ and $1/z$ around $z=\infty$.
Since there is no holomorphic one-form on the sphere the above
correlator can be determined from the singularities.
When $z \to \infty$ we use the operator
product expansion
$$ J_a(z) \Phi_j (\infty ) = - \sum_n z^{n-1} \, (J_{a,n}\Phi_j)
(\infty ), \eqn\singinft$$
and we obtain
$$\lim_{z\to \infty} \omega = -\sum_{n\geq 0} z^{n-1}
\VEV{\O_\mu (w) \Phi_i(0) (J_{a,n}\Phi_j)(\infty ) } dz \, , \eqn\singinf$$
where the $n=0$ term is included since it is singular at $\infty$.
Furthermore
$$\eqalign{
\lim_{z\to 0} \omega &= \sum_{n\geq 0}{1\over z^{n+1}}
\VEV{\O_\mu (w) (J_{a,n}\Phi_i)(0 ) \Phi_j(\infty ) } dz \cr
\lim_{z\to w} \omega &= {1\over (z-w)^2}
\VEV{(J_{a,1}\O_\mu ) (\bar w)
\Phi_i(0) \Phi_j(\infty ) } dz \, ,\cr}\eqn\singzerow$$
where in the second relation we used \opeoj . There is a
small subtlety, the term with $n=0$ in the limit when $z\to 0$ is
also singular at $z=\infty$. In fact, given the Ward identity
for the charge $J_{a,0}$ we have that
$${dz\over z} \VEV{\O_\mu (w) (J_{a,0} \Phi_i) (0) \Phi_j (\infty )}
= -{dz\over z} \VEV{\O_\mu (w) \Phi_i (0)(J_{a,0} \Phi_j) (\infty ) },
\eqn\fixsing$$
which equals the $n=0$ term in the $z\to \infty$ expansion. Thus this term
should not be included. The final result is
$$\eqalign{
\omega\, = \,& {\partial \over \partial w} \Bigl\{ {dz \over {z-w}}
\VEV{(J_{a,1}\O_\mu ) (\bar w) \Phi_i(0) \Phi_j(\infty ) } \Bigr\}\cr
{}&-\sum_{n\leq -1} {dz\over z^{n+1}}
\VEV{\O_\mu (w) \Phi_i(0) (J_{a,n}\Phi_j)(\infty ) } \cr
&+\sum_{n\geq 0}{dz\over z^{n+1}}
\VEV{\O_\mu (w) (J_{a,n}\Phi_i)(0 ) \Phi_j(\infty ) }.\cr}\eqn\ward$$

\subsection{Calculation 2.}
Now we consider another identity between operator product coefficients.
By definition we have that
$$ {1\over 2\pi} H_{\mu J_{a,n}i,j} \equiv
\VEV{\O_\mu (1) (J_{a,n}\Phi_i)(0)\Phi_j(\infty )}=
\oint {dz\over 2\pi i} z^n
\VEV{\O_\mu (1) J_a(z) \Phi_i(0) \Phi_j(\infty )} , \eqn\derfid$$
where the contour surrounds $z=0$. This contour can be replaced
by two contours surrounding $z=1$ and $z=\infty$ respectively.
Using \opeoj\ we obtain
$${1\over 2\pi} H_{\mu J_{a,n}i,j} =
-n \VEV{(J_{a,1}\O_\mu ) (1) \Phi_i(0)\Phi_j(\infty )}
-\VEV{\O_\mu (1) \Phi_i(0)(J_{a,-n}\Phi_j)(\infty )},\eqn\degif$$
and therefore we find
$$H_{\mu J_{a,n}i,j}+ H_{\mu i,J_{a,-n}j} = -n\, H_{(J_{a,1}\O_\mu )i,j}
\, \delta_{\Delta_i ,\Delta_j} ,\eqn\xxss$$
where the Kronecker delta reminds us that the operator product
coefficient $H_{(J_{a,1}\O_\mu )i,j}$ vanishes unless $\Delta_i=\Delta_j$
(recall that $J_{a,1}\O_\mu$ is purely antiholomorphic).

\subsection{Calculation 3.}
We consider one final identity. We claim that
$$ \D_{\mu\, J_{a,n}i,j} (\e ) + \D_{\mu\, i,J_{a,-n}j} (\e ) = 0 , \quad
\hbox{for} \quad n\geq 0 .\eqn\eqdiv$$
This equation follows by using \expsub\ and \xxss . We
get zero because the Kronecker deltas that arise impose conditions that
cannot be satisfied when $n\geq 0$.

\APPENDIX{B}{B: \ Computations involved in curvature}

\section{Curvature of $c$}

We must substitute the result for $F_{\mu\nu}$ in Eqn.\xhummm\ into the right
hand side of Eqn.\curvpr . Consider the first term in this right
hand side, to be called $(I)$.
The first term of $F_{\mu\nu}$ gives no contribution to $(I)$
due to antisymmetry, and the second term gives
$$\eqalign{
(I) &=
\hskip-5pt\int_{|z-y_i|>1}\hskip-8pt d^2z \sum_i
\hskip-6pt\int_{|w_i| < 1}
\hskip-10pt d^2w_i \biggl( \bra{\Sigma ;z,w_i}- \bra{\Sigma ;z}
\hbox{Sg}^{w_i} \bra{0,w_i, \infty^{*(i)}} \biggr)
\ket{\O_{[\mu } (w_i)}\ket{\O_{\nu ]}(z)}\cr
&= \hskip-5pt\int_{|z-y_i|>1}
\hskip-8pt d^2z \, \bra{\Sigma ; z} \sum_i \hskip-5pt\int_{|w_i| < 1}
\hskip-10pt d^2w_i \, \hbox{Fp}^{w_i} \bigl( \bra{0,w_i, \infty^{*(i)}}\bigr)
\ket{\O_{[\mu } (w_i)}\ket{\O_{\nu ]}(z)} .\cr }\eqn\firstpart$$
The second term in the right hand side of \curvpr , to be called $(II)$,
gives
$$\eqalign{
(II) &= \sum_i \hskip-5pt\int_{|z-y_i|<1}\hskip-10pt d^2z\,
\hbox{Fp}^{z-y_i}\hskip-10pt\int_{|w-y_i|>1}
\hskip-10pt d^2w \bra{\Sigma ;z,w}
\bigl(\ket{\O_{[\mu } (w)}\ket{\O_{\nu ]}(z)}\bigr)\cr
&+\sum_{i,j} \hskip-2pt\int_{|z-y_i|<1}\hskip-10pt d^2z\,
\hbox{Fp}^{z-y_i}\hskip-10pt\int_{|w_j|<1}
\hskip-10pt d^2w_j \biggl( \bra{\Sigma ;z,w_j}- \bra{\Sigma ;z}
\hbox{Sg}^{w_j} \bra{0,w_j, \infty^{*(j)}} \biggr)
\ket{\O_{[\mu } (w_j)}\ket{\O_{\nu ]}(z)}\cr}\eqn\secterm$$
Since $w$ is outside the unit disks,
the first term in $(II)$ can be written as
$$\int_{|w-y_i|>1}\hskip-10pt d^2w \bra{\Sigma ;w}
\sum_i \hskip-5pt\int_{|z_i|<1}\hskip-10pt d^2z_i\,\hbox{Fp}^{z_i}
\bra{0,z_i,\infty^{*(i)} }\bigl(\ket{\O_{[\mu } (w)}\ket{\O_{\nu ]}(z)}\bigr),
\eqn\djhyx$$
where we introduced the coordinates $z_i=z-y_i$. Relabeling $w\to z$ and
$z_i\to w_i$ we find that this term cancels precisely against $(I)$.
The second term in $(II)$, for the case when $i\not= j$, that is,
when $z$ and $w$ lie on different unit disks, is given by
$$\eqalign{
&\quad\sum_{i\not= j} \int_{|z_i|<1}\hskip-10pt d^2z_i\,
\hbox{Fp}^{z_i}\hskip-10pt\int_{|w_j|<1}
\hskip-10pt d^2w_j \, \bra{\Sigma ;z_i} \, \hbox{Fp}^{w_j}\,
\bra{0,w_j, \infty^{*(j)}} \bigl(
\ket{\O_{[\mu } (w_j)}\ket{\O_{\nu ]}(z_i)}\bigr)\cr
&= \bra{\Sigma} \sum_{i\not= j} \int_{|z_i|<1}\hskip-8pt d^2z_i\,
\hskip-10pt\int_{|w_j|<1}\hskip-9pt d^2w_j \, \hbox{Fp}^{z_i}
\bra{0,z_i,\infty^{*(i)}} \, \hbox{Fp}^{w_j}\,
\bra{0,w_j, \infty^{*(j)}} \bigl(
\ket{\O_{[\mu } (w_j)}\ket{\O_{\nu ]}(z_i)}\bigr) = 0,\cr}\eqn\itgoes$$
and vanishes identically because of antisymmetry.
All in all, the right hand side of \curvpr\ reduces to the contribution
arising from the second term in $(II)$ for the case when $i=j$, that
is when both punctures are in the same disk. This contribution is
$$\sum_i \int_{|z_i|<1}
\hskip-8pt d^2z_i \, \hbox{Fp}^{z_i} \hskip-8pt\int_{|w_i| < 1}
\hskip-10pt d^2w_i \biggl( \bra{\Sigma ;z_i,w_i}- \bra{\Sigma ;z_i}
\hbox{Sg}^{w_i} \bra{0,w_i, \infty^{*(i)}} \biggr)
\ket{\O_{[\mu } (z_i)}\ket{\O_{\nu ]}(w_i)}\eqn\sstlthr$$
Since both the $z_i$ and $w_i$ integrals are now restricted to
the same unit disks, we may write
$$\bra{\Sigma ;z_i,w_i}- \bra{\Sigma ;z_i}
\hbox{Sg}^{w_i} \bra{0,w_i, \infty^{*(i)}}  \quad\quad\quad$$
$$\quad\quad\quad =\bra{\Sigma}
\biggl( \bra{0 ,z_i,w_i, \infty^{*(i)}}-
\bra{0^{k},z_i,\infty^{*(i)} }
\hbox{Sg}^{w_i} \bra{0,w_i, \infty^{*k}} \biggr), \eqn\factorsigma$$
where $\bra{0 ,z_i,w_i, \infty^{*(i)}}$ is a four punctured sphere
with standard coordinates around the punctures, and $k$ is just
a label for state spaces to be contracted.
If we substitute this back into \sstlthr\ we can now identify
the curvature operator as
$$\Omega_{\mu\nu} = \int_{|z|<1}
\hskip-8pt d^2z \, \hbox{Fp}^z \hskip-8pt\int_{|w| < 1}
\hskip-8pt d^2w \biggl( \bra{0 ,z,w, \infty^*} - \bra{0^k,z,\infty^*}
\hbox{Sg}^w \bra{0,w, \infty^{*k} } \biggr)
\ket{\O_{[\mu }}\ket{\O_{\nu ]}} .\eqn\result$$

\section{Curvature in terms of OPE Coefficients}

Starting from Eqn.\result we would like to obtain an expression for
$\Omega_{\mu\nu}(c)$ in terms of the OPE coefficients. This can always be done
since all of the tensors involved can be expressed in terms of three point
functions. We know from Eqn. \result\ that
$$\Omega_{\mu\nu} = \int_{|z|<1}
\hskip-8pt d^2z \,  I(z) .\eqn\intfpz
$$
where the matrix $I$ is given by
$$
I(z) = \hbox{Fp}^z \hskip-8pt\int_{|w| < 1}
\hskip-8pt d^2w \biggl( \bra{0 ,z,w, \infty^*} - \bra{0^k,z,\infty^*}
\hbox{Sg}^w \bra{0,w, \infty^{*k} } \biggr)
\ket{\O_{[\mu }(z)}\ket{\O_{\nu ]}(w)}.\eqn\idefine$$
We  will write out the various tensors involved in the above equation in
index notation. The index representation of three point functions is obtained
from the definition of the OPE coefficients (See Eqn. \opeex\ ) as
$$\bra{\Phi^k}\,\bra{0,z,\infty^*}\ket{\Phi_i}\ket{\O_\mu} = {1\over 2\pi}
{H_{\mu
i}^{~~k} \over |z|^{2+\gamma_i-\gamma_k}}e^{-i\theta (s_i-s_k)} .\eqn\edefope
$$
Since all of the integrals we will be doing involve rotational symmetry we can
work with simpler expressions by integrating over the $\theta$ dependence
above. With the definitions $r =|z|$ and $t=|w|$ we find that the replacement
of interest is
$$\int_{|z| = r} d \theta \,\bra{\Phi^k}\bra{0,z,\infty^*}\ket{\Phi_i}
\ket{\O_\mu}={H_{\mu i}^{~~k} \over r^{2+\gamma_i-\gamma_k}}\delta_{s_i,s_k}
,\eqn\eintdefope$$

Let us begin our computation. It follows directly from the general
expression for curvature that  $\Omega_{\mu\nu i}^{~~~j} = 0$
for $\gamma_i < \gamma_j$ since $c$ is upper triangular. This means that we
need to compute $\Omega_{\mu\nu i}^{~~~j}$ only for $\gamma_i \geq \gamma_j$.
Henceforth we will restrict our attention to this case.
We write $I(z)= I_<(z) + I_{>}(z)$ where $I_<$ and $I_>$ denote
the contributions from $t < r$ and  $t > r$ respectively. For $t < r$ the
four punctured sphere can be built by sewing two three punctured spheres
in the following fashion
$$\bra{0,z,w,\infty^*}=\bra{0,w,\infty^{*k}}\bra{0^k,z,\infty^*}.\eqn\bfps$$
This imples that
$$\int_{|z| = r} d \theta \,[I_<(z)]_i^{~j}  = \delta_{s_i,s_j}\, \hbox{Fp}^z
\int_{t < r}t dt \, \sum_{s_k=s_i \atop \gamma_k > \gamma_i} {H_{[\nu
i}^{~~k} \over t^{2+\gamma_i-\gamma_k}} {H_{\mu ]k}^{~~j} \over
r^{2+\gamma_k-\gamma_j}}. \eqn\tlessr$$
Since the sums over intermediate states $k$ will always be restricted to
states such that $s_k=s_i (=s_j)$ we will drop this condition henceforth
for brevity. For $t > r$ the four-punctured sphere is built as follows
$$\bra{0 ,z,w, \infty^*} =
\bra{0,z,\infty^{*k}}\bra{0^k,w,\infty^*}.\eqn\xbfp$$
and this yields
$$\int_{|z| = r} d \theta [I_>(z)]_i^{~j} = \delta_{s_i,s_j}\, \hbox{Fp}^r
\int_{t > r}t dt  \biggl( \sum{H_{[\mu i}^{~~k} \over
r^{2+\gamma_{ik}}}{H_{\nu ]k}^{~~j} \over t^{2+\gamma_{kj}}}  +
\sum_{\gamma_k \leq\gamma_i} {H_{[\mu i}^{~~k} \over
t^{2+\gamma_{ik}}} {H_{\nu ]k}^{~~j} \over r^{2+\gamma_{kj}}}\biggr)
.  \eqn\tgreatr$$
Now we perform the integrals over $t$ in Eqns.\tlessr\ and \tgreatr\ and
take the finite part, as instructed in \idefine . We find
$$
\int_{|z| = r} d \theta\, [I_<(z)]_i^{~j}= 0,\eqn\getnothing$$
and
$$\int_{|z| = r} d \theta\, [I_>(z)]_i^{~j} = -\delta_{s_i,s_j} \biggl(
\sum  {H_{[\mu i}^{~~k} H_{\nu ]k}^{~~j} \over \gamma_{kj}}
 {1 \over r^{2+\gamma_{ik}}}
 + \sum_{ \gamma_i \geq \gamma_k} {H_{[\mu i}^{~~k} H_{\nu]
k}^{~~j}\over {\gamma_{ik}}} {1 \over
r^{2+\gamma_{kj}}}\biggr).\eqn\tgreaterr$$
Performing the integration in $r$ we obtain the final answer
$$
\Omega_{\mu\nu i}^{~~~j}  = \delta_{s_i,s_j}
\biggl( \,\sum_{\gamma_k > \gamma_i} + \sum_{\gamma_k < \gamma_j }\,\biggr)
\, {H_{[\mu i}^{~~k} H_{\nu ] k}^{~~j} \,\delta_{s_k,s_i} \over
\gamma_{kj}\, \gamma_{ki}}\, , \;\;\;\;
\hbox{for} \,\,\gamma_i \geq \gamma_j ,\eqn\curvcfin$$
and $\Omega_{\mu\nu i}^{~~~j}  = 0$, for $\gamma_i < \gamma_j$.

\refout
\end